\documentclass[preprint2,times,tighten]{aastex6}

\usepackage{amsmath,amstext}
\usepackage[figure,figure*]{hypcap}

\shorttitle{Metallicities in the low-$z$ CGM}
\shortauthors{Prochaska et al.}

\newcommand{\nwerk}{29}
\newcommand{\nmetal}{32}  
\newcommand{\vmcgm}{9.2}  
\newcommand{\emcgm}{4.3}  
\newcommand{\vfesc}{0.70 \pm 0.07}  
\newcommand{\mmetal}{1.0 \pm 5.6}  
\newcommand{\analymin}{one}
\newcommand{\nsys}{14}
\newcommand{\nqso}{13}
\newcommand{\nlow}{6}
\newcommand{\nhigh}{7}
\newcommand{\medzh}{-0.51}
\newcommand{\zhcl}{[-1.71,0.76]}
\newcommand{\syserr}{0.15\,dex}
\newcommand{\mtedge}{\tau_{912}}

\newcommand{\mtll}{\tau_{\rm LL,\lambda}}
\newcommand{\tll}{$\mtll$}
\newcommand{\msol}{M_\odot}
\newcommand{\mmhot}{M_{\rm CGM}^{\rm hot}}
\newcommand{\mhot}{$\mmhot$}
\newcommand{\mmhalo}{M_{\rm halo}}
\newcommand{\mhalo}{$\mmhalo$}
\newcommand{\mmcgm}{M_{\rm CGM}^{\rm cool}}
\newcommand{\mcgm}{$\mmcgm$}
\newcommand{\mrperp}{R_\perp}
\newcommand{\rperp}{$\mrperp$}
\newcommand{\mfecgm}{f_{\rm esc}^{\rm CGM}}
\newcommand{\fecgm}{$\mfecgm$}
\newcommand{\mftau}{f_{\tau \ge 1}}
\newcommand{\ftau}{$\mftau$}
\newcommand{\mftot}{f_{\rm esc}^{\rm T}}
\newcommand{\mmcA}{M_{\rm CGM}^{\rm Ann,i}}
\newcommand{\mcA}{$\mmcA$}
\newcommand{\ftot}{$\mftot$}

\def\perd{\;\;\; .}
\def\cmma{\;\;\; ,}

\def\ltp{\left ( \,}

\def\rtp{\, \right  ) }

\def\smm{\sum\limits}

\newcommand{\mnhi}{N_{\rm HI}}
\newcommand{\nhi}{$\mnhi$}
\newcommand{\mNH}{N_{\rm H}}
\newcommand{\NH}{$\mNH$}
\newcommand{\mbNH}{N_{\rm H,j}^{\rm best}}
\newcommand{\bNH}{$\mbNH$}

\newcommand{\cm}[1]{\, {\rm cm^{#1}}}

\newcommand{\lya}{Ly$\alpha$}

\newcommand{\N}[1]{{N({\rm #1})}}
\newcommand{\sci}[1]{{\rm \; \times \; 10^{#1}}}

\begin{document}
\title{The COS-Halos Survey: Metallicities in the Low-Redshift Circumgalactic Medium \altaffilmark{1}}
\author{J. Xavier Prochaska\altaffilmark{2}, 
Jessica K. Werk\altaffilmark{2,3},
G\'abor Worseck\altaffilmark{4}, 
Todd M. Tripp\altaffilmark{5},
Jason Tumlinson\altaffilmark{6,7},
Joseph N. Burchett\altaffilmark{5}
Andrew J. Fox\altaffilmark{6},
Michele Fumagalli\altaffilmark{8},
Nicolas Lehner\altaffilmark{9},
Molly S. Peeples\altaffilmark{6,7},
Nicolas Tejos\altaffilmark{10, 11}
}

\altaffiltext{1}{Based on observations made with the NASA/ESA Hubble Space Telescope, obtained at the Space Telescope Science Institute, which is operated by the Association of Universities for Research in Astronomy, Inc., under NASA contract NAS 5-26555. These observations are associated with programs 13033 and 11598.}
\altaffiltext{2}{University of California, Santa Cruz; xavier@ucolick.org}
\altaffiltext{3}{University of Washington, Department of Astronomy, Seattle, WA 98195}
\altaffiltext{4}{Max-Planck-Institut f\"{u}r Astronomie, K\"{o}nigstuhl 17, D-69117 Heidelberg, Germany}
\altaffiltext{5}{Department of Astronomy, University of Massachusetts, 710 North Pleasant Street, Amherst, MA 01003-9305}
\altaffiltext{6}{Space Telescope Science Institute, Baltimore, MD, 21218}
\altaffiltext{7}{Department of Physics and Astronomy, Johns Hopkins University, Baltimore, MD, 21218}
\altaffiltext{8}{Institute for Computational Cosmology and Centre for Extragalactic Astronomy, Department of Physics, Durham University, South Road, Durham, DH1 3LE, UK}
\altaffiltext{9}{Center of Astrophysics, Department of Physics, University of Notre Dame, 225 Nieuwland Science Hall, Notre Dame, IN 46556}
\altaffiltext{10}{Millennium Institute of Astrophysics, Santiago, Chile}
\altaffiltext{11}{Instituto de Astrof\'isica, Pontificia Universidad Cat\'olica de
Chile, Vicu\~na Mackenna 4860, Santiago, Chile}

\begin{abstract}
We analyze new far-ultraviolet spectra of \nqso~quasars from the 
$z\sim 0.2$ COS-Halos survey that cover the \ion{H}{1} Lyman limit of 
\nsys~circumgalactic medium (CGM) systems.
These data yield precise estimates or more constraining limits
than previous COS-Halos
measurements on the \ion{H}{1} column densities \nhi.
We then apply a Monte-Carlo Markov Chain approach on 32~systems
from COS-Halos to estimate
the metallicity of the cool ($T\sim\,10^4$\,K) CGM gas
that gives rise to low-ionization state metal lines, under the assumption
of photoionization equilibrium with the extragalactic UV background.
The principle results are:
(1) the CGM of field $L^*$ galaxies exhibits a declining \ion{H}{1}
surface density with impact parameter \rperp\ (at $>99.5\%$ confidence),
(2) the transmission of ionizing radiation through CGM
gas alone is $70\pm7\%$;
(3) the metallicity distribution function of the cool CGM is
unimodal with a median
of $10^{\medzh}\,Z_\odot$ and a 95\%\ interval
$\approx1/50~Z_\odot$ to $>3~Z_\odot$.
The incidence of metal poor ($<1/100\,Z_\odot$) gas is low,
implying any such gas discovered along quasar sightlines is
typically unrelated to $L^*$ galaxies;
(4) we find an unexpected increase in gas metallicity
with declining \nhi\ (at $>99.9\%$ confidence)
and, therefore, also with increasing \rperp. 
The high metallicity at large radii implies early enrichment.
(5) A non-parametric estimate of the cool CGM gas mass is
$\mmcgm=(\vmcgm\pm\emcgm)\sci{10}\msol$, which together
with new mass estimates for the hot CGM may resolve
the galactic missing baryons problem. 
Future analyses of halo gas should focus on the underlying
astrophysics governing the CGM, 
rather than processes that simply expel the medium from the halo.
\end{abstract}


\section{Introduction}
\label{sec:intro}


Both the conceptualization and discovery of the circumgalactic 
medium (CGM) was based on the 
observation of heavy elements (e.g.\ \ion{Mg}{2}, \ion{C}{4}, \ion{O}{6}) 
along quasar sightlines \citep[e.g.][]{bs69,bergeron86,trippetal00,clw01,pwc+06}.
As larger surveys and datasets were compiled, it became clear that
the present-day CGM accounts for the majority if not all of the detected
metal absorption \citep{ctp+10,pwc+11,bordoloi14,lehner+15}.
Consequently, this medium is a major reservoir of heavy elements
with a mass rivaling and possibly exceeding that within
galaxies \citep{ttw+11,werk+14,peeples+14}.

Given the diffuse and highly ionized nature of the CGM, 
its metals must have originated within one or more
galaxies and have been transported to the $\sim$100 kpc
distances where we observe them. 
A number of metal transport mechanisms
have been proposed, including galactic winds, AGN feedback, accretion, tidal stripping, and ram pressure 
\citep[e.g.][]{veilleux05,putman12,os13b}.
Most of these processes depend sensitively on, and possibly govern, 
basic galaxy properties such as stellar mass, star formation rates, and chemical enrichment.  
Gas metallicities provide critical clues to the action of these processes.  
For example, a high metallicity may indicate that the CGM
is polluted by higher mass, chemically-enriched systems.
In contrast, a very low metallicity may indicate IGM accretion
and/or the by-products of lower mass, satellite galaxies
\citep{lht+13}. It is plausible that both high and low-metallicity 
gas coexists in halos from a mixture of ongoing accretion and 
feedback. If so, the balance may shift with galaxy mass or 
other properties in ways that reveal the relative importance 
of the accretion and feedback mechanisms. 

Because we use ions of heavy elements to diagnose the physical conditions in CGM gas, its metallicity also bears on its inferred total mass as traced by its \ion{H}{1} content. Even if the \ion{H}{1} column density (\nhi) is well constrained, it must be corrected for ionization to derive total surface densities and 
then integrated to estimate the gaseous halo mass. These ionization corrections are derived from the observed metal lines. Most galaxy-selected studies to date \citep[][hereafter W14]{pwc+11,stocke13,borthakur+15,werk+14} have used small, heterogeneous samples dominated by systems bearing large \nhi\ uncertainties caused by saturation in the Lyman series lines that yield lower limits to \nhi\ $\sim 10^{15-16}$ cm$^{-2}$. Sightlines penetrating the inner CGM ($\mrperp < 100$\,kpc), where \ion{H}{1} column densities are likely higher than this, are particularly affected. 
This was especially the case for the COS-Halos survey \citep{ttw+11,werk12a,tumlinson+13}
which analyzed the CGM of $z \sim 0.2$, field $L^*$ galaxies at impact
parameters $\mrperp < 150$\,kpc.
Indeed, our own previous analysis of the COS-Halos survey 
included $\sim 20$~systems with lower limits
to the \nhi\ values based on \ion{H}{1} Lyman series analysis \citep{tumlinson+13}. 
It is important to obtain precise \nhi\ measurements to fully understand the nature of CGM gas.  For example, with access to higher 
Lyman series lines that precisely constrain \nhi, \cite{rlh+11} 
show that a  saturated \lya\ absorber at an impact parameter of 37\,kpc 
has a much lower metallicity than its host galaxy 
and therefore may be an example of cool gas accretion.
Recognizing this limitation to the measurement of CGM gas masses and metallicities, we carried out new observations with the Cosmic Origins Spectrograph (COS) on the {\it Hubble Space Telescope} (HST) to cover the \ion{H}{1} Lyman Limit (LL) of \nsys\ systems. This manuscript describes those observations and the new results that follow.

Section~\ref{sec:obs} describes the new {\it HST}/COS
observations and Section~\ref{sec:NHI} presents the new
\nhi\ analysis.  In Section~\ref{sec:metala} we perform a new
metallicity analysis of the COS-Halos survey
using Monte Carlo Markov Chain (MCMC) techniques and
Section~\ref{sec:discuss} discusses the primary results.
We assume the WMAP9 cosmology \citep{wmap09}
and report proper distances unless otherwise specified.
All of the measurements presented here are available online
through the {\tt pyigm\footnote{https://github.com/pyigm/pyigm}}
repository.  

\begin{deluxetable}{lccccccccc}
\tablewidth{0pc}
\tablecaption{OBSERVATIONS\label{tab:obs}}
\tablehead{\colhead{Quasar}  & \colhead{$z_{\rm em}$}
& \colhead{Config.} & \colhead{$t_{\rm eff}^a$ (s)}
}
\startdata
SDSSJ091029.75+101413.6 & 0.462 & G140L & 6301\\
SDSSJ094331.61+053131.4 & 0.564 & G140L & 6520\\
SDSSJ095000.73+483129.3 & 0.589 & G130M & 9953\\
SDSSJ101622.60+470643.3 & 0.822 & G130M & 9962\\
SDSSJ113327.78+032719.1 & 0.524 & G130M & 7945\\
SDSSJ115758.72-002220.8 & 0.260 & G140L & 6109\\
SDSSJ123335.07+475800.4 & 0.382 & G130M & 8178\\
SDSSJ124154.02+572107.3 & 0.583 & G130M & 8005\\
SDSSJ132222.68+464535.2 & 0.374 & G130M & 8177\\
SDSSJ133045.15+281321.4 & 0.417 & G140L & 5922\\
SDSSJ134251.60-005345.3 & 0.326 & G130M & 9237\\
SDSSJ141910.20+420746.9 & 0.874 & G140L & 7333\\
SDSSJ155504.39+362848.0 & 0.714 & G140L & 6943\\
\hline
\enddata
\tablenotetext{a}{Median effective exposure time.}
\end{deluxetable}

\section{Observations and Data Processing}
\label{sec:obs}

We observed \nqso\ of the COS-Halos sightlines using the COS G140L/1280 setting for \nlow\ quasars and the G130M/1222 setting for \nhigh\ quasars (Cycle 20, Program 13033, PI Tumlinson). These two settings were chosen to optimize the short wavelength coverage of the new spectroscopy, extending down the range of the existing COS-Halos data (Program 11598, PI Tumlinson) 
to $\sim 1000$ \AA. Prior to these observations, we had observed and fully analyzed the targeted absorbers as part of the main COS-Halos survey, so that we were able to select the  G140L/1280 setting for systems at $z < 0.2$ and G130M/1222 for $z > 0.2$ (Table~\ref{tab:obs}). The former covers shorter observed-frame wavelengths while the latter offers higher spectral resolution. The observations occurred between 2012 December and 2013 June when COS was in its second lifetime position (LP2). 

All of the data were reduced with the CALCOS pipeline (v2.21) using the COS calibration files as of 2014 December. The reduction pipeline settings were customized to use rectangular boxcar extraction windows of 25 pixels and 35 pixels for G140L and G130M spectra, respectively. Detector pulse heights were restricted to $2\le\mathrm{PHA}\le 15$ on both detector segments to preserve all source counts while minimizing the detector dark current.
These choices preserve spectrophotometric accuracy, while minimizing the background and maximizing data quality. 

After extraction, individual sub-exposures were further processed with software developed for the analysis of COS spectra in the low-count (i.e.\ Poisson) and low-flux ($f_\lambda\la 10^{-17}$\,erg\,cm$^{-2}$\,s$^{-1}$\,\AA$^{-1}$) regime \citep{worseck+16}.
Briefly, this software estimates the COS pulse-height-restricted dark current in the science aperture using contemporaneous dark calibration exposures obtained at the same detector voltage and similar space-weather conditions within $\pm 1.5$ months around the date of observation, and coadds subexposures in count space while flagging detector blemishes. The post-processed dark current is accurate to a few percent \citep{worseck+16}, which is crucial for our measurements of nearly saturated Lyman continuum absorption ($\log\mnhi\simeq 17.8$).

In G140L spectra scattered geocoronal Ly$\alpha$ emission can be significant \citep{worseck+16}, but this background component could not be directly estimated, since geocoronal Ly$\alpha$ is not covered in the G140L 1280\,\AA\ setup.
Based on our analysis of deep G130M data of \ion{He}{2}-transparent quasars \citep{shull10,syphers14,worseck+16}
we consider scattered light negligible for G130M spectra at the wavelengths of interest (i.e.\ $\lambda<$1200\,\AA). 
Diffuse Galactic and extragalactic
sky emission was not subtracted, as it is much lower than the dark current and scattered light \citep[only 4--10\% of the total background;][]{worseck+16}. 
Geocoronal oxygen and nitrogen emission was effectively eliminated by considering shadow 
(orbital night-only)
data in the affected wavelength ranges if available. 
Residual geocoronal emission was flagged after visual inspection.

\begin{figure}
\begin{center}
\includegraphics[width=3.5in]{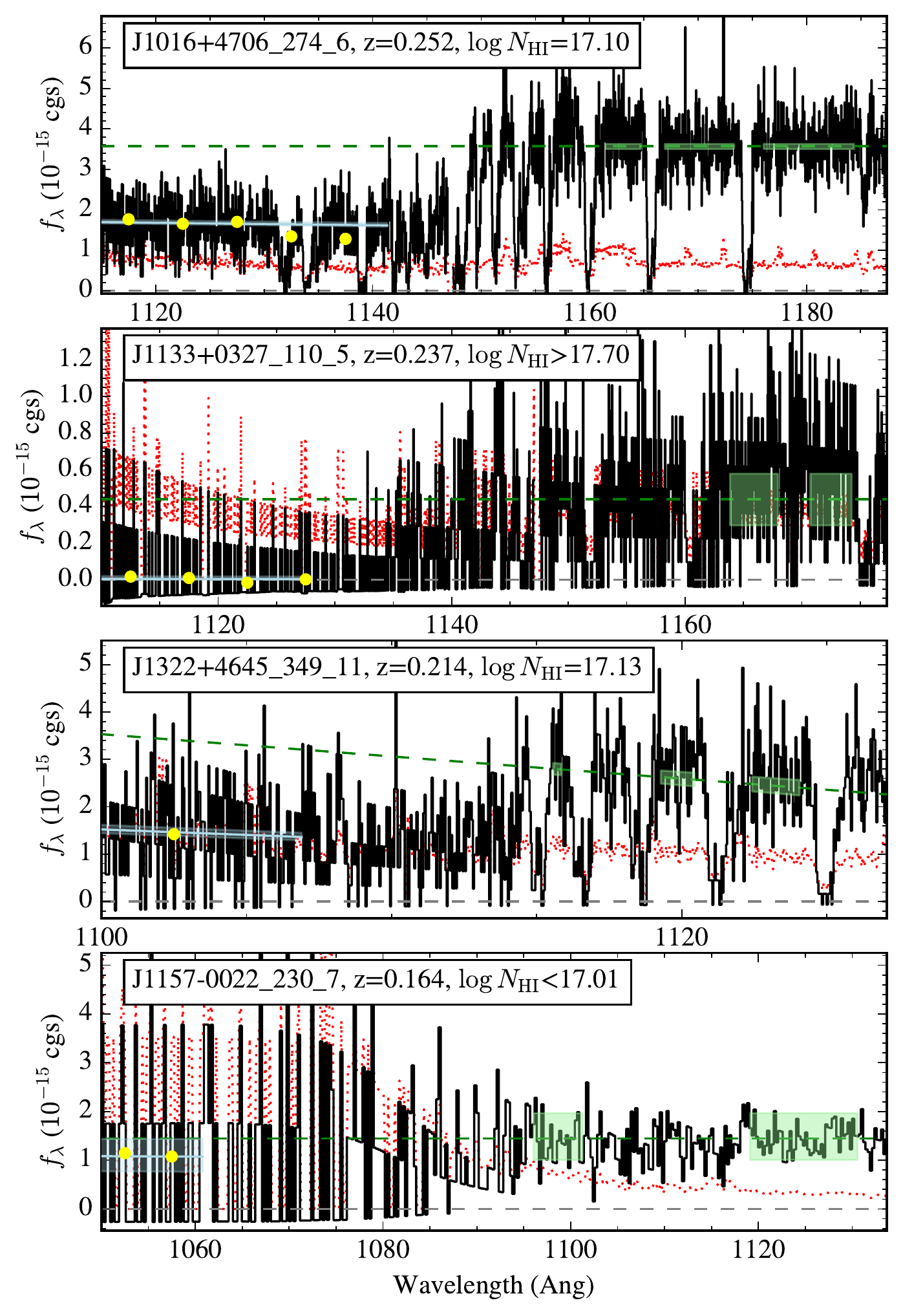}
\caption{
Zoom-in figures of the Lyman limits of four representative
CGM systems from the sample.  The black line is the HST/COS
spectra with corresponding $1\sigma$ error array (red; dotted).
The green dashed line is the best-fit continuum model $C_\lambda$
and the green shaded regions show the spectral regions used to
fit $C_\lambda$.  The blue line shows the model flux within the
Lyman limit region and the shaded blue regions indicate the spectral regions used to estimate \tll.  
The yellow dots show the mean flux in 5\AA\ bins.
These four examples show a range of \nhi\ values and spectral quality characteristic of the full sample, 
including an upper limit (J1157--0022\_230\_7) and
a lower limit (J1133+0327\_110\_5).
}
\label{fig:NHI_models}
\end{center}
\end{figure}

For final analysis the G140L spectra were binned by a factor of 3 and the G130M spectra by a factor of 4, resulting in a sampling of approximately 2 pixels per resolution element at COS Lifetime Position 2 (G140L: resolving power $\lambda/\Delta\lambda\simeq 2\,000$ at 1150\,\AA, dispersion $\simeq 0.24$ \AA\,pixel$^{-1}$; G130M: $\lambda/\Delta\lambda\simeq 15\,000$ at 1150\,\AA, dispersion $\simeq 0.04$ \AA\,pixel$^{-1}$). For display purposes we computed an approximate $1\sigma$ error array following \citet{worseck+16}, but use the correct asymmetric Poisson error throughout our analysis. Examples of the fully reduced COS spectra are shown in Figure~\ref{fig:NHI_models}, zoomed in on the regions where the Lyman limit absorption occurs.  These examples illustrate the range of data quality. 
The remainder of the sample is shown in the Appendix.

\begin{figure*}
\includegraphics[width=7in]{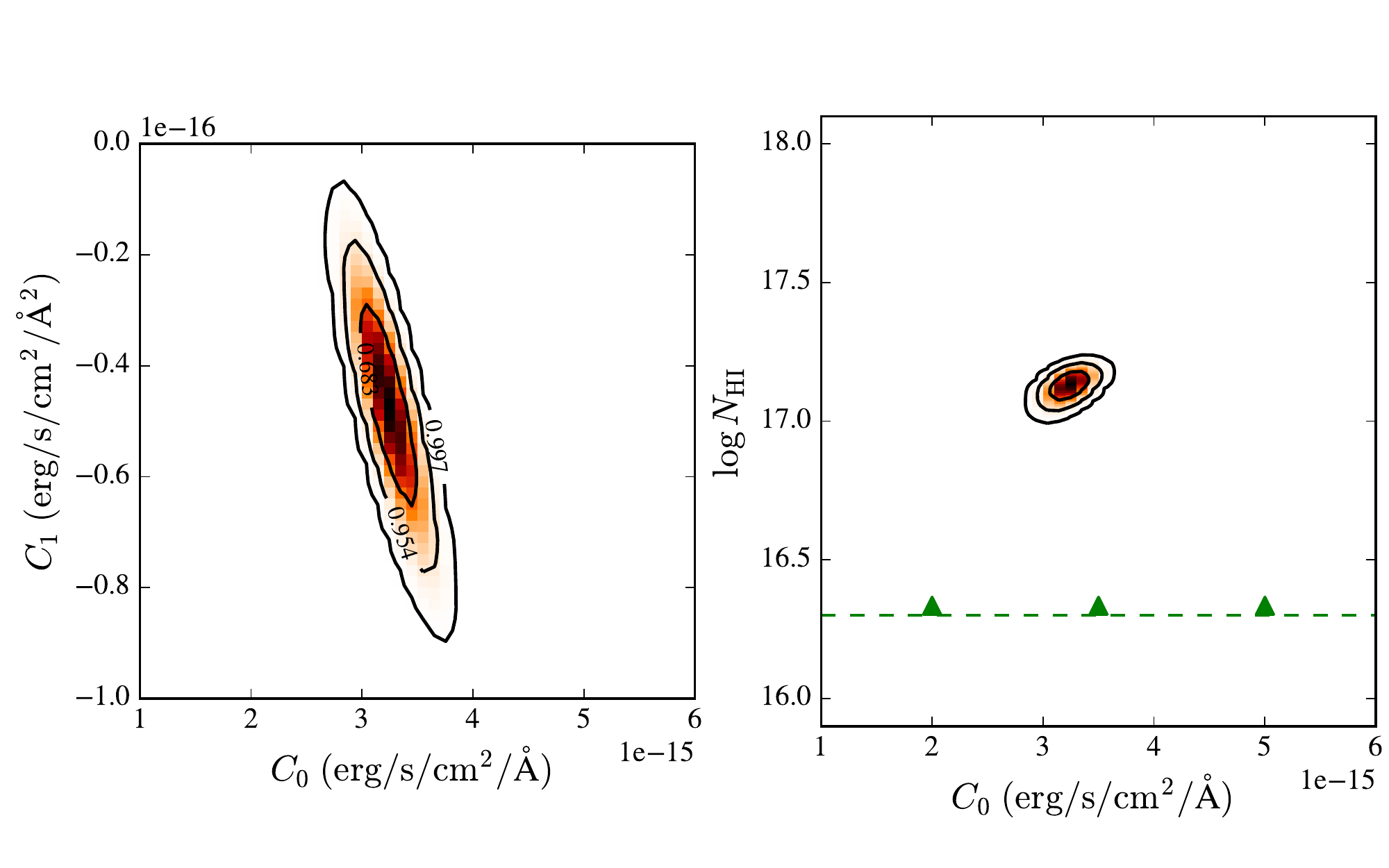}
\caption{
Model constraints on the continuum $C_\lambda$ of quasar
J1322+4645 with $C_\lambda = C_0 + C_1(\lambda-911)$\AA\
(left)
and on the \ion{H}{1} column density \nhi\ of the
CGM associated to the galaxy 349\_11 at $z=0.214$ (right).
The contours, from inner to outer, indicate the 0.683,
0.954, and 0.997 confidence limits.
This system shows an \ion{H}{1} Lyman limit opacity near
unity which yields a tight constraint on \nhi.
For comparison, the dashed line shows the previous lower
limit on \nhi\ based on Lyman series analysis,
and the previous upper
limit was $\mnhi < 10^{19} \cm{-2}$ \citep{tumlinson+13}. 
}
\label{fig:contours}
\end{figure*}

\section{\nhi\ Analysis}
\label{sec:NHI}

Our program was designed to provide \nhi\ estimates for \nsys\ systems from the
COS-Halos sample through the observations of \nqso\ quasars.
These new spectra cover the \ion{H}{1} Lyman limit of each system,
enabling an estimate of the 
continuum opacity: 

\begin{equation}
\mtll \approx \tau_{912} \ltp \frac{\lambda}{912\mathrm{\AA}} \rtp^{2.75}
\end{equation}
with $\tau_{912}$ the optical depth at 
energy $h\nu_{912}=1$\,Ryd.
From this opacity, one recovers 
a direct estimate of the total \nhi\ for the system,

\begin{equation}
\mnhi =  \tau_{912} / \sigma(1 \, \rm Ryd) \cmma
\label{eqn:NHI}
\end{equation}
where $\sigma(1 \, \rm Ryd) \approx 6.34 \sci{-18} \cm{2}$ 
is the \ion{H}{1} photoionization cross-section for 1\,Ryd photons.
In the following, we adopt the \cite{verner96} parameterization
of \tll\ which gives an accurate representation of the quantum
mechanical derivation.

The measurement of \tll\ requires an estimate of 
the quasar continuum $C_\lambda$, 
\begin{equation}
\mtll = -\ln(f_\lambda/C_\lambda) \;\;\; .
\end{equation}
We extrapolate a model for $C_\lambda$ based on
the observed flux just redward of the Lyman limit.
The observed, attenuated quasar flux $f_\lambda$ is also partially 
absorbed by lines associated to the CGM,
the Galactic ISM, and other absorption systems along the sightline.
This absorption is generally weak
and one can identify spectral regions that are likely
unabsorbed.

We employed two approaches to estimating $C_\lambda$, depending on the absorption redshift and the spectral S/N. For each system we adopt one of these two approaches:
(i) a linear fit to $C_\lambda$ using select regions of
the unabsorbed data, $C_\lambda = C_0 + C_1(\lambda-911$\AA);
and
(ii) a constant value, $C_\lambda=C_0$ fitted to select
regions. We adopt the former for data of higher quality based on 
inspection near the Lyman limit. Examples of our adopted continuum placements are shown in Figure~\ref{fig:NHI_models}. The remaining systems are presented in the Appendix.

For J0943+0531, the LL occurs just blueward of the $\sim 18$\AA\ gap 
between the COS FUV detector segments, which precludes a direct estimate of $C_\lambda$ near the break. We set $C_\lambda=C_0$ based on the flux measured redward of the gap (at $\lambda \approx 1350$\AA) and adopt a large uncertainty.
This system has negligible 
LL absorption and we recover only a conservative upper limit to \nhi\ which is 
relatively insensitive to $C_\lambda$.

After setting the unabsorbed continuum regions, we performed a maximum likelihood analysis to estimate \nhi. The full model flux $M_\lambda$ consists of a continuum $C_\lambda$, parameterized
by $C_0$ and/or $C_1$, attenuated by the Lyman limit opacity \tll\
set by the free parameter \nhi\ given in 
Equation~\ref{eqn:NHI}:
\begin{equation}
M_\lambda = C_\lambda \, \exp(-\mtll) \perd
\end{equation}
From this model flux, the average number of model counts per pixel
$\mu_i$ was estimated using the known sensitivity function for the instrument and the effective exposure time 
$t_{\rm eff}$ (Table~\ref{tab:obs}) of each observation.  
Additionally, we included an estimate of the background
counts following \cite{worseck+16}. We assumed a Poisson deviate for the counts in the LL region and Gaussian statistics for the continuum. Formally, the Poisson deviate for the observed counts $m_i$ in each pixel of the analysis region is $P_i(m_i;\mu_i)$.
The likelihood function follows simply as $\mathcal{L} = \Pi_i P_i$.

We calculated the maximum likelihood $\mathcal{L}$ for
a large grid covering the allowed space for the continuum parameters and \nhi. The best values of these three parameters are taken at the maximum $\mathcal{L}$. Errors are estimated by integrating over the grid to assess the cumulative probability. Figure~\ref{fig:contours} shows the results for a well-modeled
system (J1322+4645\_349\_11\footnote{
Throughout this paper, we adopt the COS-Halos notation
for naming CGM systems, composed of the quasar name, then
the position angle (deg) and angular offset (deg) from
the quasar sightline.}); 
it describes the constraints on the parameters
and also their correlation.  Analysis of the LL
yields only limits to \nhi\ when the opacity is much higher or lower than unity. 
In these cases, we report one-sided 95\%~confidence limits for \nhi\ and
rely on the Lyman series analysis to further refine the value.

Figure~\ref{fig:NHI_models} shows the best-fit models
of four systems overlaid on the spectra near the LL (the remainder of systems are shown in the Appendix). Table~\ref{tab:NHI_fits} lists the spectral regions used for the Lyman limit and continuum analyses, the best-fit values, and the $68\%$ confidence interval for \nhi. We have also revisited the Lyman series analysis for
the systems with only lower limits on \nhi. In nearly every case, we have set an upper limit from the absence of damping wings in the
\lya\ line. These are also provided in the table, and we consider all values in this range equally likely (i.e. we adopt a flat prior). 

\clearpage

\begin{deluxetable*}{lccccccccccccc}
\tablewidth{0pc}
\tablecaption{\nhi\ MODELS AND MEASUREMENTS\label{tab:NHI_fits}}
\tabletypesize{\small}
\tablehead{\colhead{Galaxy} & \colhead{z} & \colhead{flg$_C^a$} & \colhead{$\lambda_C^b$} 
& \colhead{$C_0$} & \colhead{$\sigma(C_0)$} 
& \colhead{$C_1$} & \colhead{$\sigma(C_1)$} 
& \colhead{$\lambda_\tau$} & \colhead{$\log \mnhi^c$} 
& \colhead{C.L.$^d$} 
\\ 
 & & & (\AA) & (10$^{-15}$) & & (10$^{-17}$) &  } 
\startdata 
J0910+1014\_242\_34 & 0.264& 2& [1157.30,1159.95]& 0.64 & 0.02& & & [1120.0,1150.0]& $ 16.58$& 16.51,16.62\\ 
&&& [1162.68,1165.80]\\ 
J0910+1014\_34\_46 & 0.143&0&&&&&&& $ 17.25$& 16.00,18.50\\ 
J0943+0531\_106\_34 & 0.228& 2& [1127.04,1132.19]& 1.96 & 0.10& & & [1090.0,1119.0]& $< 16.24$& \\ 
&&& [1136.35,1140.84]\\ 
J0943+0531\_227\_19 & 0.353& 3&& 2.01 & & & & [1120.0,1150.0]& $< 16.65$& \\ 
J0950+4831\_177\_27 & 0.212& 1& [1116.38,1117.77]& 5.71 & 0.15& $-3.02$ & 0.49& [1095.0,1104.0]& $> 17.91$& \\ 
&&&&&&&&& 18.20 & 17.90,18.50\\ 
J1009+0713\_170\_9 & 0.356&0&&&&&&& $ 18.50$& 18.00,19.00\\ 
J1009+0713\_204\_17 & 0.228&0&&&&&&& $ 17.50$& 16.50,18.50\\ 
J1016+4706\_274\_6 & 0.252& 2& [1161.42,1164.69]& 3.57 & 0.06& & & [1115.0,1130.0]& $ 17.10$& 17.08,17.11\\ 
&&& [1166.92,1173.34]\\ 
&&& [1176.01,1184.38]\\ 
J1016+4706\_359\_16 & 0.166&0&&&&&&& $ 17.50$& 16.50,18.50\\ 
J1112+3539\_236\_14 & 0.247&0&&&&&&& $ 16.70$& 15.80,17.60\\ 
J1133+0327\_110\_5 & 0.237& 2& [1163.83,1167.97]& 0.44 & 0.14& & & [1110.0,1120.0]& $> 17.70$& \\ 
&&&&&&&&& 18.60 & 18.54,18.66\\ 
J1133+0327\_164\_21 & 0.154&0&&&&&&& $ 16.90$& 15.80,18.00\\ 
J1157-0022\_230\_7 & 0.164& 2& [1095.88,1101.43]& 1.45 & 0.48& & & [1050.0,1063.0]& $< 17.01$& \\ 
&&& [1118.98,1130.69]\\ 
J1233+4758\_94\_38 & 0.222& 2& [1130.63,1131.36]& 4.31 & 0.15& & & [1106.0,1111.0]& $ 16.74$& 16.69,16.77\\ 
&&& [1138.68,1142.35]\\ 
J1241+5721\_199\_6 & 0.205& 1& [1110.63,1111.52]& 3.71 & 0.21& $-3.76$ & 0.57& [1091.0,1098.0]& $> 17.83$& \\ 
&&&&&&&&& 18.15 & 17.80,18.50\\ 
J1322+4645\_349\_11 & 0.214& 1& [1115.55,1115.86]& 3.24 & 0.10& $-4.69$ & 0.71& [1100.0,1106.0]& $ 17.14$& 17.10,17.16\\ 
&&& [1119.25,1120.45]\\ 
&&& [1122.41,1124.07]\\ 
J1330+2813\_289\_28 & 0.192& 2& [1102.32,1109.35]& 2.02 & 0.29& & & [1070.0,1086.0]& $ 17.03$& 16.88,17.11\\ 
J1342-0053\_157\_10 & 0.227& 1& [1131.02,1132.28]& 5.29 & 0.21& $-1.71$ & 0.53& [1105.0,1117.5]& $> 18.04$& \\ 
&&&&&&&&& 18.50 & 18.00,19.00\\ 
J1419+4207\_132\_30 & 0.179& 2& [1094.70,1099.28]& 3.00 & 0.24& & & [1045.0,1074.0]& $ 16.63$& 16.33,16.72\\ 
&&& [1106.57,1110.78]\\ 
J1514+3619\_287\_14 & 0.212&0&&&&&&& $ 17.50$& 16.50,18.50\\ 
J1555+3628\_88\_11 & 0.189& 2& [1100.62,1105.08]& 0.60 & 0.12& & & [1057.0,1083.5]& $ 17.17$& 16.91,17.30\\ 
&&& [1108.25,1113.18]\\ 
\hline 
\enddata 
\tablenotetext{a}{Flag describing the continuum method applied: 0=Analysis based only on Lyman series lines; 1=Linear fit; 2=Constant fit; 3=Continuum imposed by hand.}
\tablenotetext{b}{Wavelength interval used to fit the quasar continuum redward of the Lyman limit.}
\tablenotetext{c}{$\log \mnhi$ value.  If reported as a limit, this corresponds to a one-sided $95\%$ confidence interval.
For lower limits, we report a second \nhi\ value and interval that is bounded at the high end by analysis of the \lya\ line.}
\tablenotetext{d}{Confidence interval on $\log \mnhi$. If the interval exceeds 0.5~dex, one should assume a uniform prior.
Otherwise, the interval covers 68\% of an approximately Gaussian prior.}
\tablecomments{Units for $C_0$ and $C_1$ are erg/s/cm$^2$/\AA\ and erg/s/cm$^2$/\AA$^2$ respecitvely.}
\end{deluxetable*} 


\begin{figure}
\begin{center}
\includegraphics[width=3.5in]{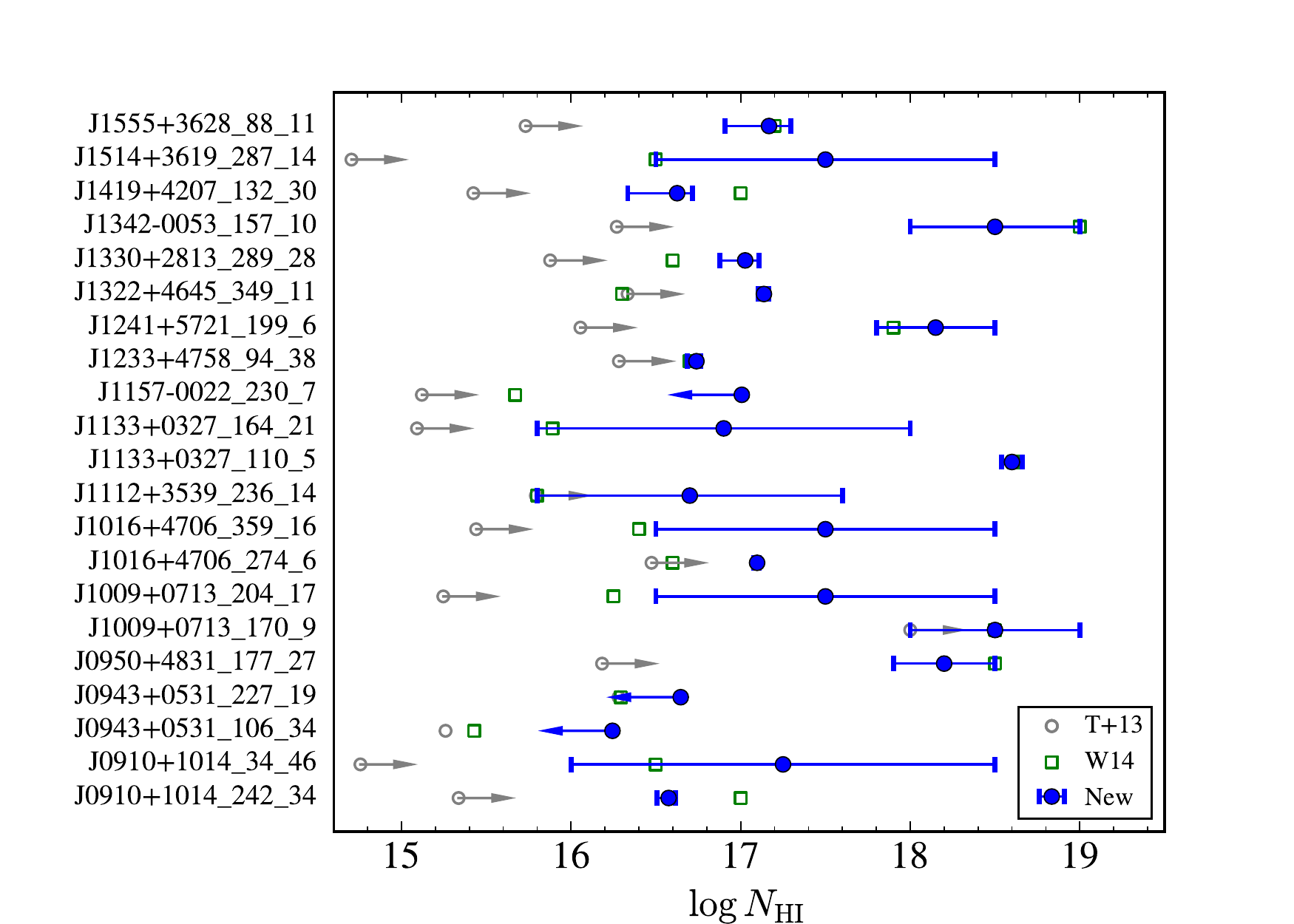}
\end{center}
\caption{
Measurements of \nhi\ including upper and lower limits
from the analysis in this paper (blue)
compared against previous estimates from \ion{H}{1}
Lyman series analysis \citep[][ gray circles and 
green squares, respectively]{tumlinson+13,werk+14}.
The new values are in general agreement with the 
previous estimates and limits.
}
\label{fig:diffNHI}
\end{figure}

Figure~\ref{fig:diffNHI} compares these new \nhi\ measurements with our previous estimates. The \cite{tumlinson+13} measurements (grey circles) were conservatively derived from analysis of the \ion{H}{1} Lyman series while the W14 estimates (green squares) included a prior on the gas metallicity, requiring sub-solar values. The new \nhi\ measurements (blue circles with errors) exceed prior estimates from the Lyman series, or impose a lower limit consistent with the previous measurement.

\begin{figure}
\begin{center}
\includegraphics[width=3.5in]{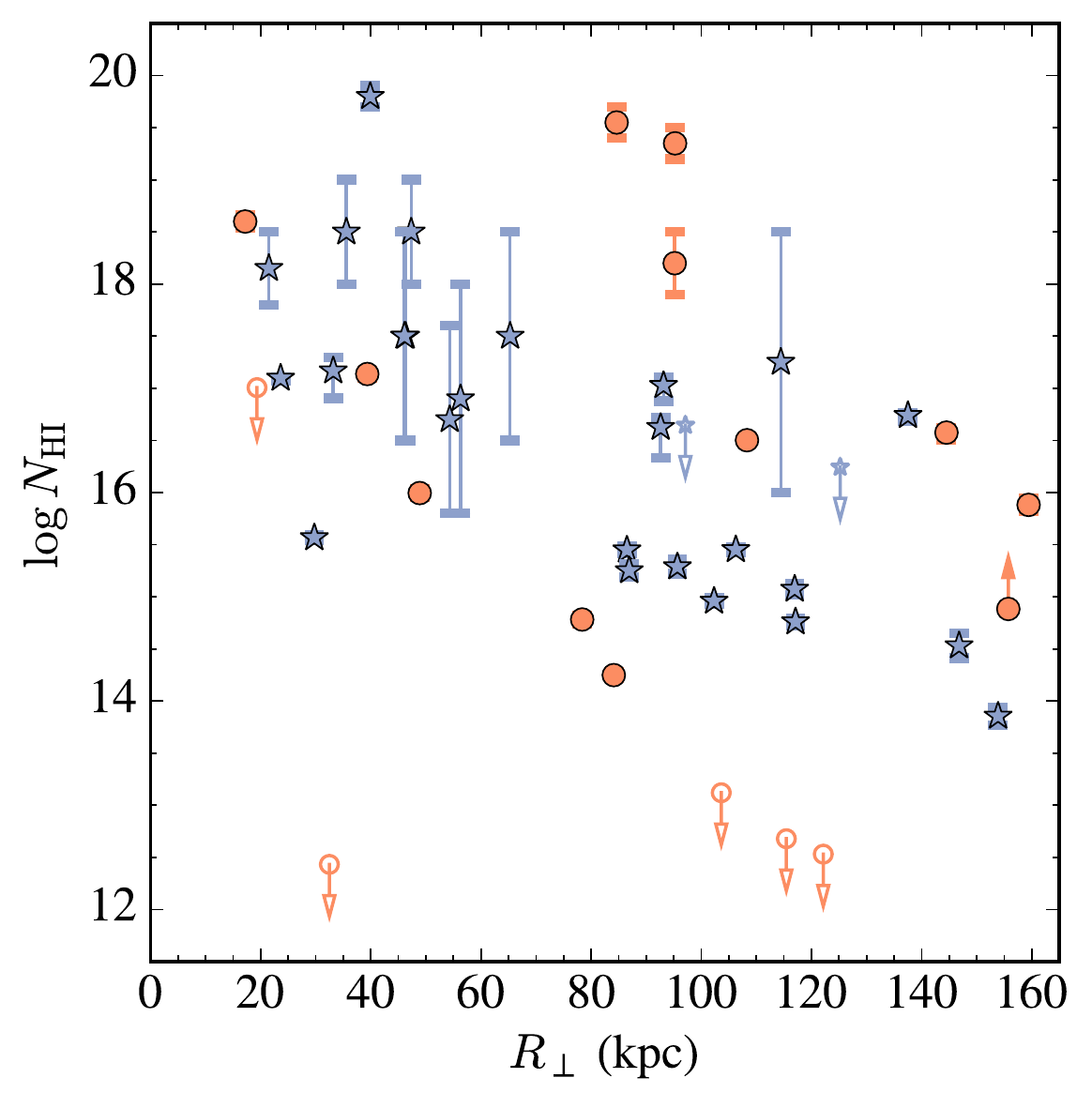}
\end{center}
\caption{
\nhi\ values for the COS-Halos survey
versus the projected impact
parameter \rperp\ to the target galaxy.  
The measurement of each CGM system
is coded by the specific SFR (sSFR) 
such that red circles indicate 
a sSFR $<$ 10$^{-11}$ M$_{\odot}$ yr$^{-1}$, 
while blue stars represent galaxies with sSFR higher than this limit.  Open symbols
indicate non-detections in \ion{H}{1} and error bars (typically $<0.1$\,dex) are overplotted on detections. Note the high incidence
of optically thick gas at $\mrperp < 60$\,kpc and the overall
trend to lower \nhi\ at higher \rperp.  The null hypothesis
of no correlation is ruled out at $>99.6\%$ confidence.
}
\label{fig:NHI_vs_rho}
\end{figure}

Figure~\ref{fig:NHI_vs_rho} presents the updated
\nhi\ distribution for the COS-Halos sample as a function
of impact parameter and color-coded for the 
target galaxy's star-formation rate \citep{werk12a}.
We find a strong anti-correlation between \nhi\ and \rperp\ across
the full sample;  a Kendall Tau correlation test including censored data carried out with the ASURV package\footnote{The Astronomy SURVival analysis (Rev. 1.3) package can be downloaded via http://astrostatistics.psu.edu/statcodes/asurv \citep{asurv85}} 
rules out the null hypothesis at $>99.6\%$ confidence.
This suggests a higher characteristic hydrogen density
$n_{\rm H}$ closer to the galaxy.

As an aside, we emphasize that the COS-Halos sample
exhibits a high incidence of optically thick gas
($\mnhi \ge 10^{17} \cm{-2}$) for sightlines penetrating
within $\approx 60$\,kpc of a field $L^*$ galaxy. This implies that long sightlines selected for high redshift and/or bright FUV magnitudes are somewhat biased against the `inner' CGM for luminous galaxies at $z > 0.3 - 0.5$ because the strong LL absorption in the inner CGM severely suppresses the FUV flux.





\section{Metallicity Analysis}
\label{sec:metala}

In this section, we reexamine the metal enrichment of the cool CGM in the COS-Halos sample with two key advances over previous work. First, the new \nhi\ measurements greatly improve the precision of  the gas metallicity estimates. Second, we adopt a new methodology for constraining the ionization state based on the techniques described in \cite{fumagalli+16a} (see also \citet{crighton+15}).


\subsection{Methodology}

We have used Monte Carlo Markov Chain (MCMC)
techniques to compare a grid of plane-parallel, photoionization models
parameterized by the gas density $n_{\rm H}$, \ion{H}{1} column density
\nhi, and metallicity [Z/H] against the observed ionic column 
densities of low ionization state 
metal species (e.g. Si$^{+}$, Si$^{++}$, N$^{+}$).
A detailed description of this analysis used to estimate gas metallicities
for the COS-Halos sample is provided in the Appendix. 
From the MCMC chains, we derive probability distribution functions (PDFs)
for the model parameters that describe the physical state of the 
absorbing gas.  These provide a more quantitative estimation
of the statistical uncertainties than our previous analysis. 
Table~\ref{tab:ionz_models}
summarizes the main results for the \nmetal\ 
systems analyzed.

\begin{deluxetable*}{lccccccccc}
\tablewidth{0pc}
\tablecaption{SUMMARY OF PHOTOIONIZATION MODELING \label{tab:ionz_models}}
\tabletypesize{\scriptsize}
\tablehead{\colhead{Galaxy} & \colhead{m$_{\rm ion}^a$}
& \colhead{$f^b$} & \colhead{$\log N_{\rm HI,min}^{\rm prior}$}
& \colhead{$\log N_{\rm HI,max}^{\rm prior}$}
& \colhead{$\log \mnhi$} & \colhead{$\log n_{\rm H}$}
& \colhead{[Z/H]} 
} 
\startdata 
J0401-0540\_67\_24& 2& 0& 15.42 & 15.48& 15.34, 15.39, 15.45& -3.89, -3.56, -3.30& -0.27, -0.10, 0.15\\ 
J0803+4332\_306\_20& 1& 0& 14.74 & 14.82& 14.61, 14.69, 14.80& -3.82, -3.02, -2.69& -0.48, 0.06, 0.80\\ 
J0910+1014\_242\_34& 5& 0& 16.52 & 16.63& 16.42, 16.52, 16.63& -2.77, -2.59, -2.42& -0.25, -0.17, -0.04\\ 
J0910+1014\_34\_46& 4& -3& 16.00 & 18.50& 17.45, 17.71, 17.94& -4.03, -3.80, -3.53& -1.77, -1.62, -1.42\\ 
J0914+2823\_41\_27& 1& 0& 15.42 & 15.48& 15.34, 15.40, 15.45& -3.81, -3.36, -3.07& -0.66, -0.44, -0.14\\ 
J0925+4004\_196\_22& 8& 0& 19.40 & 19.70& 19.39, 19.51, 19.65& -4.42, -4.17, -3.88& -0.95, -0.81, -0.66\\ 
J0928+6025\_110\_35& 8& 0& 19.20 & 19.50& 19.13, 19.30, 19.47& -3.11, -2.96, -2.85& -0.40, -0.15, 0.14\\ 
J0943+0531\_106\_34& 1& 0& 15.50 & 16.57& 15.53, 15.94, 16.39& -3.83, -3.36, -2.92& -1.28, -0.70, -0.13\\ 
J0950+4831\_177\_27& 7& -3& 17.90 & 18.50& 18.03, 18.20, 18.37& -2.98, -2.80, -2.63& -1.01, -0.91, -0.77\\ 
J1009+0713\_170\_9& 7& -3& 18.00 & 19.00& 18.34, 18.62, 18.84& -2.76, -2.53, -2.39& -0.92, -0.76, -0.61\\ 
J1009+0713\_204\_17& 3& -3& 16.50 & 18.50& 17.13, 17.26, 17.39& -4.02, -3.81, -3.63& -1.19, -1.03, -0.83\\ 
J1016+4706\_274\_6& 6& 0& 17.08 & 17.11& 17.00, 17.05, 17.10& -3.21, -3.08, -2.98& -0.40, -0.35, -0.30\\ 
J1016+4706\_359\_16& 5& -3& 16.50 & 18.50& 16.80, 17.74, 18.24& -3.77, -3.39, -3.14& -1.55, -1.23, -0.43\\ 
J1112+3539\_236\_14& 2& -3& 15.80 & 17.60& 16.20, 16.68, 17.20& -2.81, -2.66, -2.53& -1.46, -0.93, -0.44\\ 
J1133+0327\_110\_5& 6& 0& 18.54 & 18.66& 18.40, 18.52, 18.59& -3.40, -3.20, -2.99& -1.39, -1.27, -1.19\\ 
J1220+3853\_225\_38& 2& 0& 15.82 & 15.94& 15.71, 15.82, 15.90& -4.28, -3.91, -3.60& 0.27, 0.67, 1.10\\ 
J1233+4758\_94\_38& 5& 0& 16.70 & 16.78& 16.68, 16.74, 16.80& -3.11, -2.98, -2.84& -0.38, -0.29, -0.18\\ 
J1233-0031\_168\_7& 2& 0& 15.54 & 15.59& 15.42, 15.52, 15.59& -3.80, -3.46, -3.19& -0.23, -0.00, 0.32\\ 
J1241+5721\_199\_6& 9& -3& 17.80 & 18.50& 18.25, 18.37, 18.44& -3.28, -3.20, -3.13& -0.71, -0.65, -0.59\\ 
J1241+5721\_208\_27& 2& 0& 15.22 & 15.35& 15.09, 15.21, 15.29& -3.41, -3.32, -3.19& 0.15, 0.26, 0.41\\ 
J1245+3356\_236\_36& 1& 0& 14.72 & 14.80& 14.60, 14.66, 14.72& -3.84, -3.00, -2.64& -0.57, 0.03, 0.81\\ 
J1322+4645\_349\_11& 5& 0& 17.11 & 17.17& 17.00, 17.05, 17.10& -3.10, -2.97, -2.84& -0.82, -0.70, -0.62\\ 
J1330+2813\_289\_28& 6& 0& 16.91 & 17.14& 16.88, 17.01, 17.12& -2.61, -2.50, -2.41& -0.58, -0.50, -0.41\\ 
J1342-0053\_157\_10& 9& -3& 18.00 & 19.00& 18.67, 18.82, 18.89& -2.81, -2.75, -2.70& -0.27, -0.20, -0.14\\ 
J1419+4207\_132\_30& 6& 0& 16.43 & 16.82& 16.69, 16.89, 17.07& -2.94, -2.82, -2.70& -0.69, -0.54, -0.42\\ 
J1435+3604\_126\_21& 1& 0& 15.19 & 15.31& 15.07, 15.17, 15.28& -3.90, -3.58, -3.34& -0.10, 0.14, 0.47\\ 
J1435+3604\_68\_12& 7& 0& 19.70 & 19.90& 19.59, 19.73, 19.82& -4.26, -3.76, -3.18& -1.45, -1.31, -1.18\\ 
J1514+3619\_287\_14& 2& -3& 16.50 & 18.50& 17.20, 17.51, 17.84& -2.63, -2.51, -2.42& -1.27, -1.04, -0.75\\ 
J1550+4001\_197\_23& 4& 0& 16.48 & 16.52& 16.40, 16.45, 16.50& -2.80, -2.75, -2.70& -0.40, -0.35, -0.30\\ 
J1555+3628\_88\_11& 6& 0& 16.97 & 17.36& 17.17, 17.31, 17.46& -3.23, -3.07, -2.93& -0.96, -0.82, -0.71\\ 
J2345-0059\_356\_12& 3& 0& 15.96 & 16.03& 15.82, 15.92, 16.03& -3.68, -3.52, -3.36& -0.02, 0.11, 0.27\\ 
\hline 
\enddata 
\tablenotetext{a}{Number of positive detections constraining the model.}
\tablenotetext{b}{Flag indicating \nhi\ treatment: 0=Gaussian; -3=Uniform.}
\tablecomments{The following systems had insufficient data constraints for an ionization analysis: J0226+0015\_268\_22, J0935+0204\_15\_28, J0943+0531\_216\_61, J0943+0531\_227\_19, J1133+0327\_164\_21, J1157-0022\_230\_7, J1342-0053\_77\_10, J1437+5045\_317\_38, J1445+3428\_232\_33, J1550+4001\_97\_33, J1617+0638\_253\_39, J1619+3342\_113\_40, J2257+1340\_270\_40.} 
\tablecomments{The $\log \mnhi, \log n_{\rm H}$ and [Z/H] values are 
represent the 68\% c.l. interval and median of the MCMC PDFs.}
\end{deluxetable*}

\begin{figure}
\begin{center}
\includegraphics[width=3.5in]{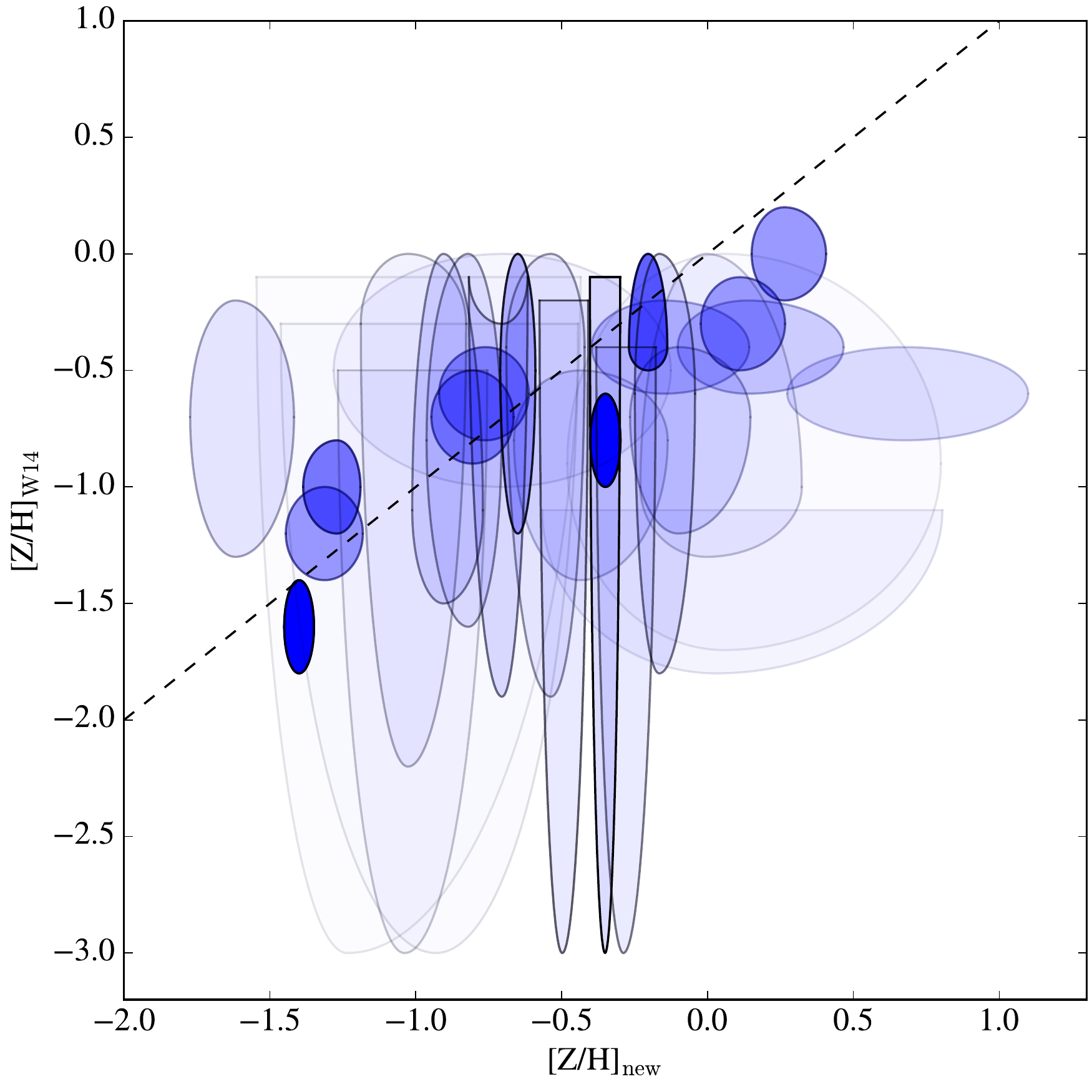}
\end{center}
\caption{
Comparison of the metallicity estimates from this work against
the values reported in W14.  The plotted ellipses have asymmetric
semi-major and semi-minor axes to mimic the asymmetries in the
PDFs for [Z/H] and/or limits to the value.
Furthermore, the shading indicates the precision in [Z/H]
in our new analysis (darker is more precise).  
Overall, there is good agreement
between the two sets of measurements which is expected
given both studies assumed photoionization modeling and
nearly the same set of ionic column densities.
The only significant difference is that the new sample
of measurements extends to [Z/H]~$> 0$ as we have relaxed
the prior adopted by W14 restricting the values to solar or lower.
The dashed lines denotes a 1-to-1 correspondence between the
two sets of measurements.
}
\label{fig:ZH_vs_W14}
\end{figure}

\subsection{Metallicity of the CGM for $L^*$ Galaxies}
\label{sec:metal}

\begin{figure}
\begin{center}
\includegraphics[width=3.5in]{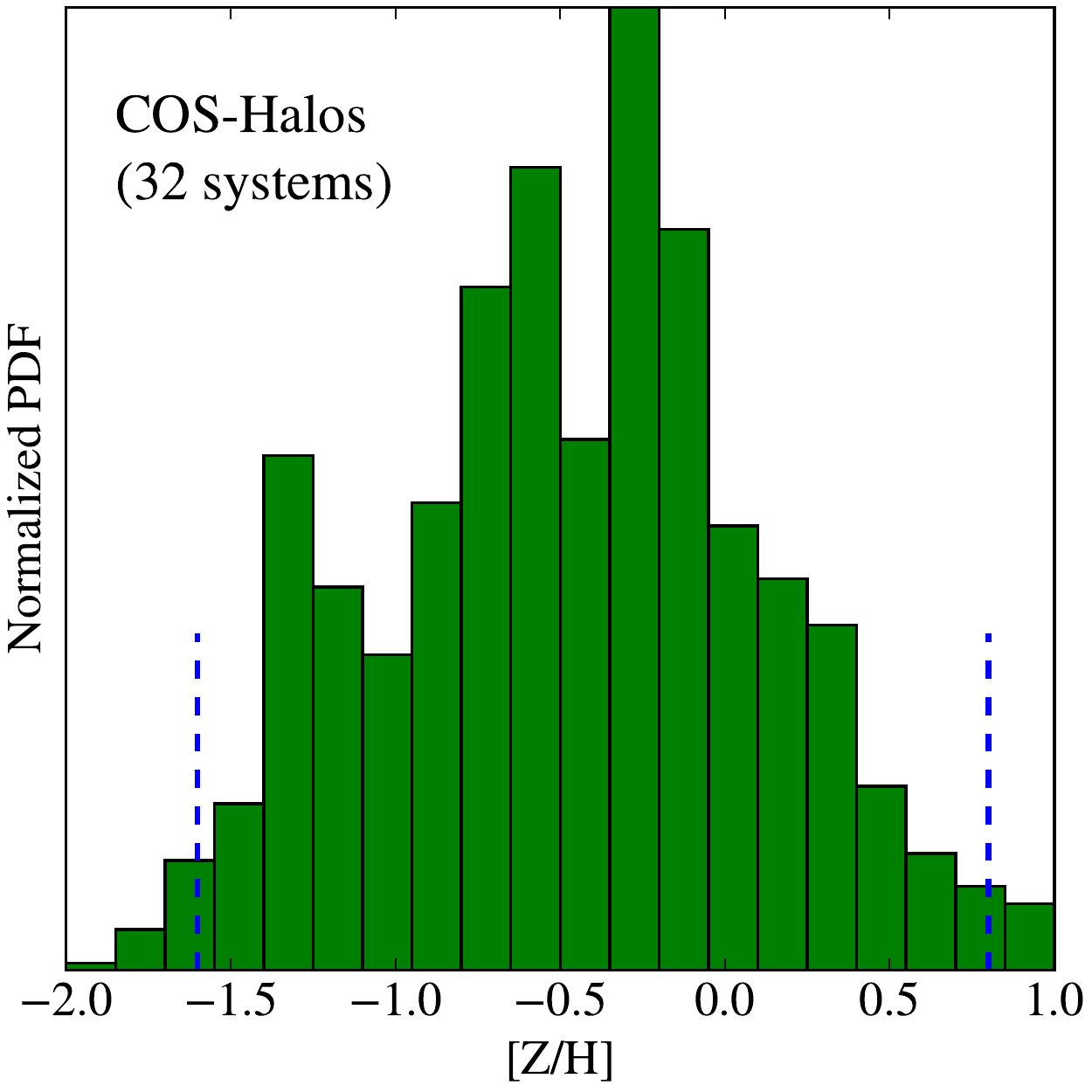}
\end{center}
\caption{
Integrated metallicity PDF for the 32\,systems from the
COS-Halos dataset with at least one positive detection of
a low or intermediate ionic state of a heavy element.
This PDF has a median value [Z/H]$_{\rm median} = \medzh$\,dex
and a 95\%\ c.l. of $\zhcl$, as marked by the blue dashed
lines.
The data is well-described by a unimodal distribution.
}
\label{fig:fullPDF}
\end{figure}

\begin{figure*}
\begin{center}
\includegraphics[width=6.5in]{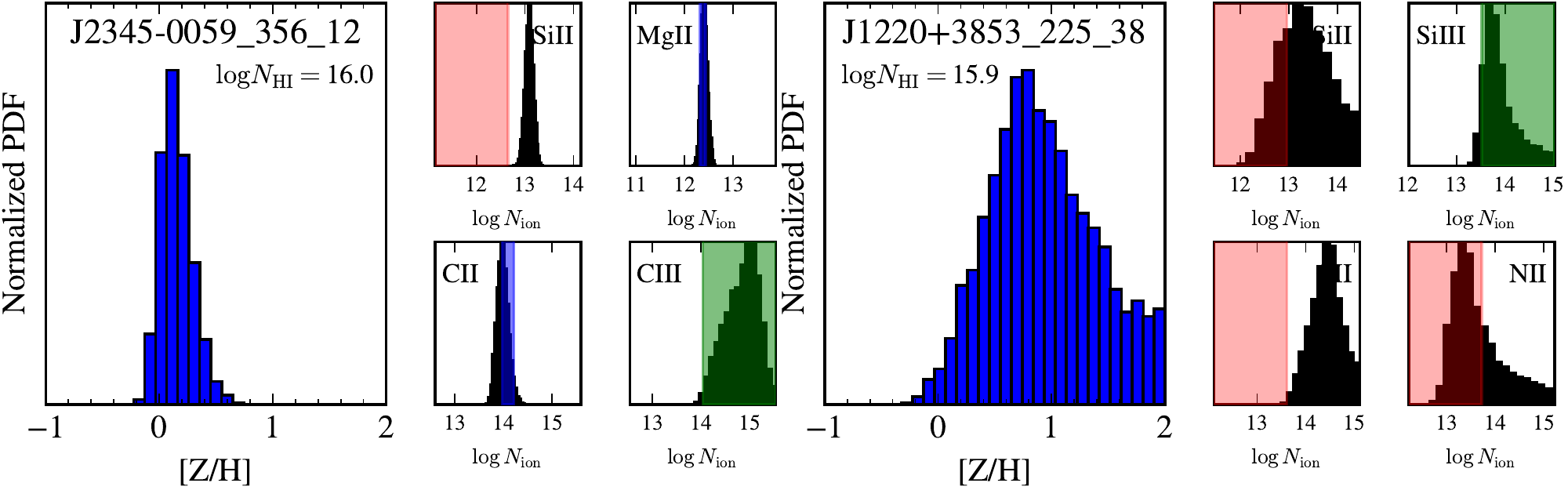}
\end{center}
\caption{ 
Metallicity PDFs for two systems where 95\%\ of the PDF exceeds 0.5 solar metallicity.  Also shown are several
of the ionic constraints for each sightline, compared 
against the model PDFs of $N_{\rm ion}$.
Shading is the same as in Figure~\ref{fig:residuals}.
}
\label{fig:super}
\end{figure*}

From the MCMC analysis, we have generated a metallicity
PDF for \nmetal\ of the COS-Halos systems with at least
one positive detection of a lower ionization state.
Figure~\ref{fig:ZH_vs_W14} compares the metallicity
measurements for the systems overlapping with W14\footnote{
Note that for CGM system J0914+2823\_41\_27, a typo in W14
reported the wrong best metallicity.  
It should have been reported as $-0.8$\,dex.}.
In general, there is good agreement between the two analyses.
This is expected given that each analysis adopted very similar observational constraints and 
assumed photoionization equilibrium. The MCMC analysis generally yields a smaller uncertainty than those reported in W14, for several reasons: 
   (a) the more precise measurements of \nhi\ from the new COS data;
  (b) a conservative approach to uncertainty estimates in W14;
 and
  (c) overly optimistic uncertainty estimates from the MCMC analysis.
On the latter point, we adopt a minimum systematic uncertainty of 0.3\,dex in metallicities due to the over-simplifying assumptions of our photoionization models \citep[e.g.\ co-spatial gas with a 
constant density;][]{haislmaier+16,wotta+16}.

Figure~\ref{fig:fullPDF} shows the combined metallicity PDF of the COS-Halos survey restricted to systems with at least
\analymin\ detected metal transition. The median gas metallicity is high, [Z/H]$_{\rm median} = -0.51$\,dex
and the 95\% interval is broad,
spanning from $\approx 1/50$ solar to 
$>$3x solar metallicity.  

We may conclude that the CGM of field $L^*$ galaxies is generally enriched above $\sim 10$\% solar. The substantial scatter in these inferred metallicities could
come from a range in the mean metallicity of the halos, from varying metallicities within
each halo, or both. We discuss these results further in Section~\ref{sec:enrich}.

\subsection{Super-solar CGM Gas}
\label{sec:super_solar}

W14 adopted a prior on the gas metallicity that restricted
[Z/H]~$\le 0$, i.e.\ to not exceed solar metallicity.  This choice
was somewhat arbitrary and was primarily motivated by the large 
\nhi\ uncertainties in a set of systems with saturated 
\ion{H}{1} Lyman series absorption. In the current analysis,
we allow [Z/H] values up to 100x solar to assess 
the incidence of super-solar metallicities.


Figure~\ref{fig:super} shows the ion constraints for two
of the four 
systems that exceed 1/2 solar metallicity at 95\%\ confidence in the MCMC analysis.  
This subset of high metallicity systems is heterogeneous 
in terms of data quality and observational constraints but all have $\mnhi < 10^{16} \cm{-2}$. The combination of low \nhi\ with the positive detection of one or more ions  drives the metallicity to a high value. 

Of course, the estimated [Z/H] values require significant
ionization corrections. Figure~\ref{fig:IC} of the Appendix shows the corrections required to convert an observed $\N{Si^{++}}/\mnhi$ ratio to a [Si/H] abundance for photoionization models with a wide range of \nhi\ and $n_{\rm H}$ values.  The figure reveals that the smallest correction 
is $\approx +2.4$\,dex and occurs at a very 
low gas density (i.e.\ a very high ionization parameter).  
For the two systems presented in Figure~\ref{fig:super}
with a Si$^{++}$ detection,
we measure $\log(\N{Si^{++}}/\mnhi) > -2.55$\,dex yielding
[Si/H]~$> -0.2$\,dex on the assumption 
of photoionization equilibrium. 
We note that similar results apply for collisional ionization
equilibrium (CIE).  Using the calculations of \cite{gs07},
the smallest ionization correction is +2.2\,dex.
Presently, we have no reason to assert that these lower 
\nhi\ systems are out of ionization equilibrium.
Furthermore, the few cases which exhibit multiple ionization
states are well-modeled by the simple equilibrium models.
Nevertheless, we caution that low density gas may not be
in strict ionization balance \citep[e.g.][]{gs07}.

In the full sample, 15\%\ of the systems have 90\% of their metallicity PDFs above solar, while 25\% of the sample has 50\% of their PDFs above solar. This implies high enrichment levels at large radii from the central galaxy. We conclude that at least a subset of the CGM surrounding field $L^*$ galaxies 
has a super-solar metallicity \citep[see also][]{tmp+11,mtw+13}.

\begin{figure*}
\begin{center}
\includegraphics[width=6.5in]{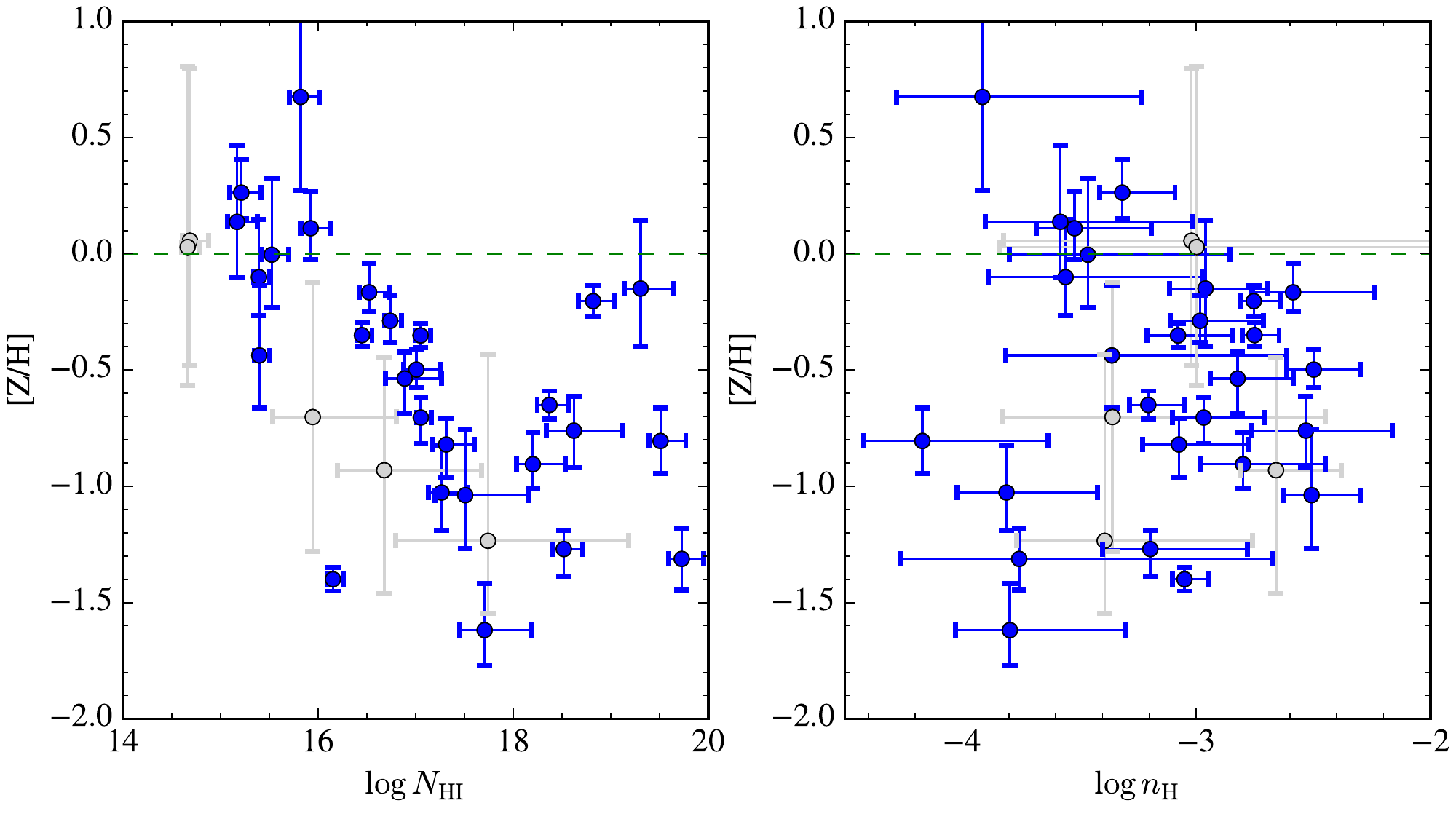}
\end{center}
\caption{
Comparison of intrinsic properties against CGM metallicity.
For the \ion{H}{1} column density (left-hand panel), there is an apparent
anti-correlation with $>99.99\%$ statistical significance.  However, [Z/H] does not correlate with the derived gas volume densities (right-hand panel).
Grey points indicate systems with very poorly constrained values.
}
\label{fig:intrinsic}
\end{figure*}

\subsection{Intrinsic Correlations}

In Figure~\ref{fig:intrinsic} we present the median
[Z/H] values and the 68\%\ confidence intervals
for the PDFs of each CGM system against several intrinsic
properties of the CGM.  From the figure, it is
apparent that there is a strong anti-correlation 
between the measured \nhi\ values and [Z/H].
A Pearson's correlation test on the plotted
values rules out the null hypothesis at $>99.99\%$ confidence.
This is driven by the approximately solar metallicity
systems with $\mnhi < 10^{16} \cm{-2}$
(see also Figure~\ref{fig:super}),
the decrease in [Z/H] with \nhi\ for systems
having $\mnhi \approx 10^{17} \cm{-2}$,
and the rarity of [Z/H]~$\approx 0$ values at high \nhi.

\begin{figure}
\begin{center}
\includegraphics[width=3.5in]{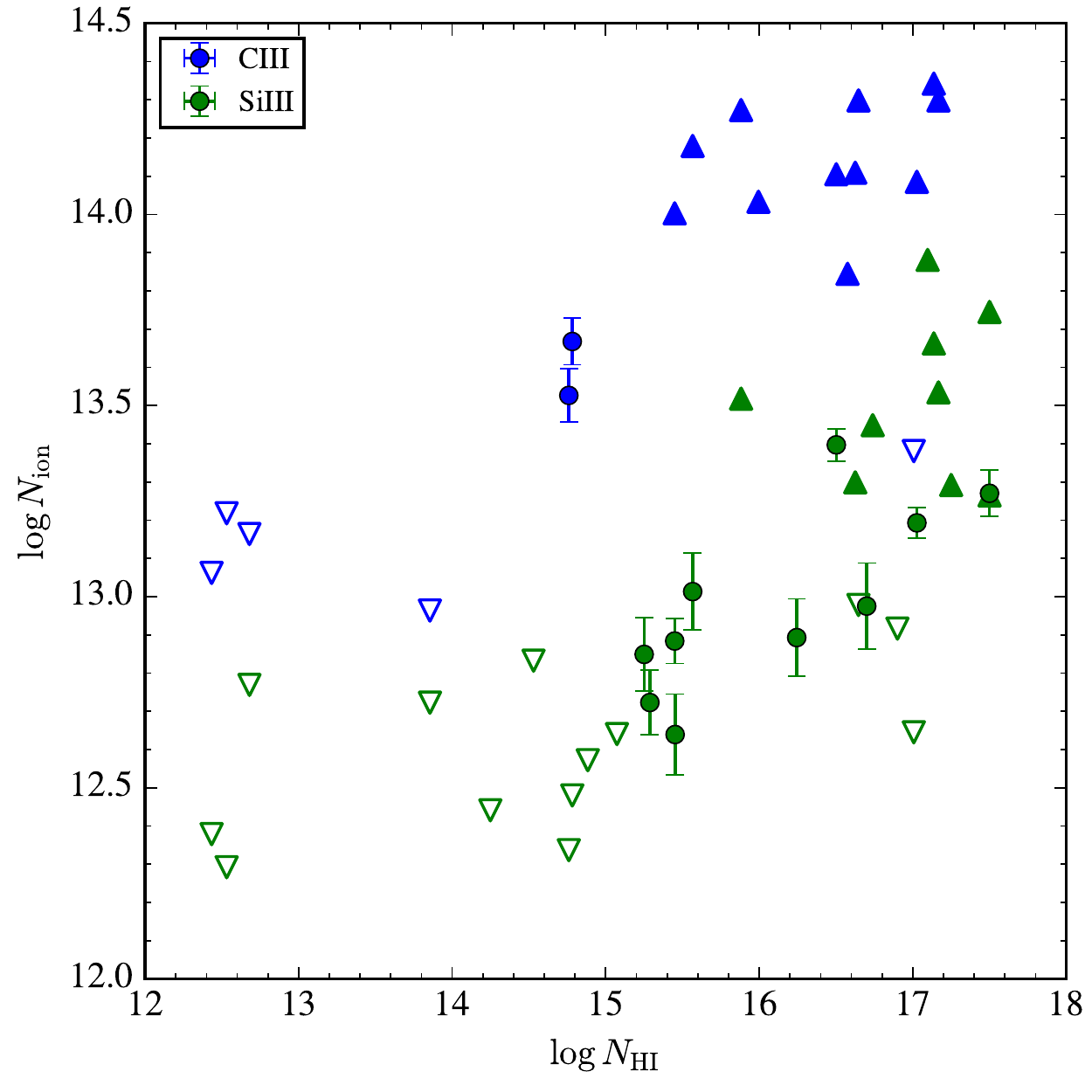}
\end{center}
\caption{
Scatter plot of Si$^{++}$ and C$^{++}$ column densities
vs.\ \nhi\ for the COS-Halos sample.
At $\mnhi > 10^{14.5} \cm{-2}$, nearly every system
exhibits a positive detection of one of these two ions.
The lower \nhi\ systems, meanwhile, have limits to 
$N_{\rm ion}$ consistent with the overall trend.
Therefore, these limits have little constraining power
on the gas metallicity.
}
\label{fig:si3c3}
\end{figure}

Before proceeding, we consider each of these points more 
carefully. First, the apparent decline in [Z/H]
 for $\mnhi = 10^{16.5-18} \cm{-2}$ could be caused by uncertainties in these \nhi\ values combined with the fact that 
[Z/H] is inversely proportional to \nhi.
However, the systems with $\mnhi \approx 10^{18} \cm{-2}$
have PDFs with values toward the low end of their allowed
\nhi\ range, which gives {\it higher} [Z/H] values.
Second, the low incidence of solar metallicity
at high \nhi\ is subject to significant sample variance.
Figure \ref{fig:intrinsic} shows that 2 of 7 systems with $\mnhi > 10^{18} \cm{-2}$
have [Z/H]~$> -0.5$\,dex. Adopting binomial statistics, the rate of high metallicity
is 0.285 with a 60\%\ uncertainty (i.e.\ a 100\%\ incidence
is nearly allowed).

We also consider whether the preponderance of 
high metallicity values at lower \nhi\ is a selection effect introduced by our requirement for at least one positive detection of a heavy element transition to perform the metallicity analysis. For low \nhi\ and limited S/N in the data, this cut prefers high metallicities. We assess this possible selection
bias as follows. Figure~\ref{fig:si3c3} plots the ionic column densities for Si$^{++}$ and C$^{++}$ for the full COS-Halos
sample against their \nhi\ values.  At 
$\mnhi \lesssim 10^{14.5} \cm{-2}$, there are no detections
and these systems may be ignored in this discussion.
At $\mnhi > 10^{15} \cm{-2}$, the detection rate
is $> 90\%$ implying no selection bias.

\begin{figure*}
\begin{center}
\includegraphics[width=6.3in]{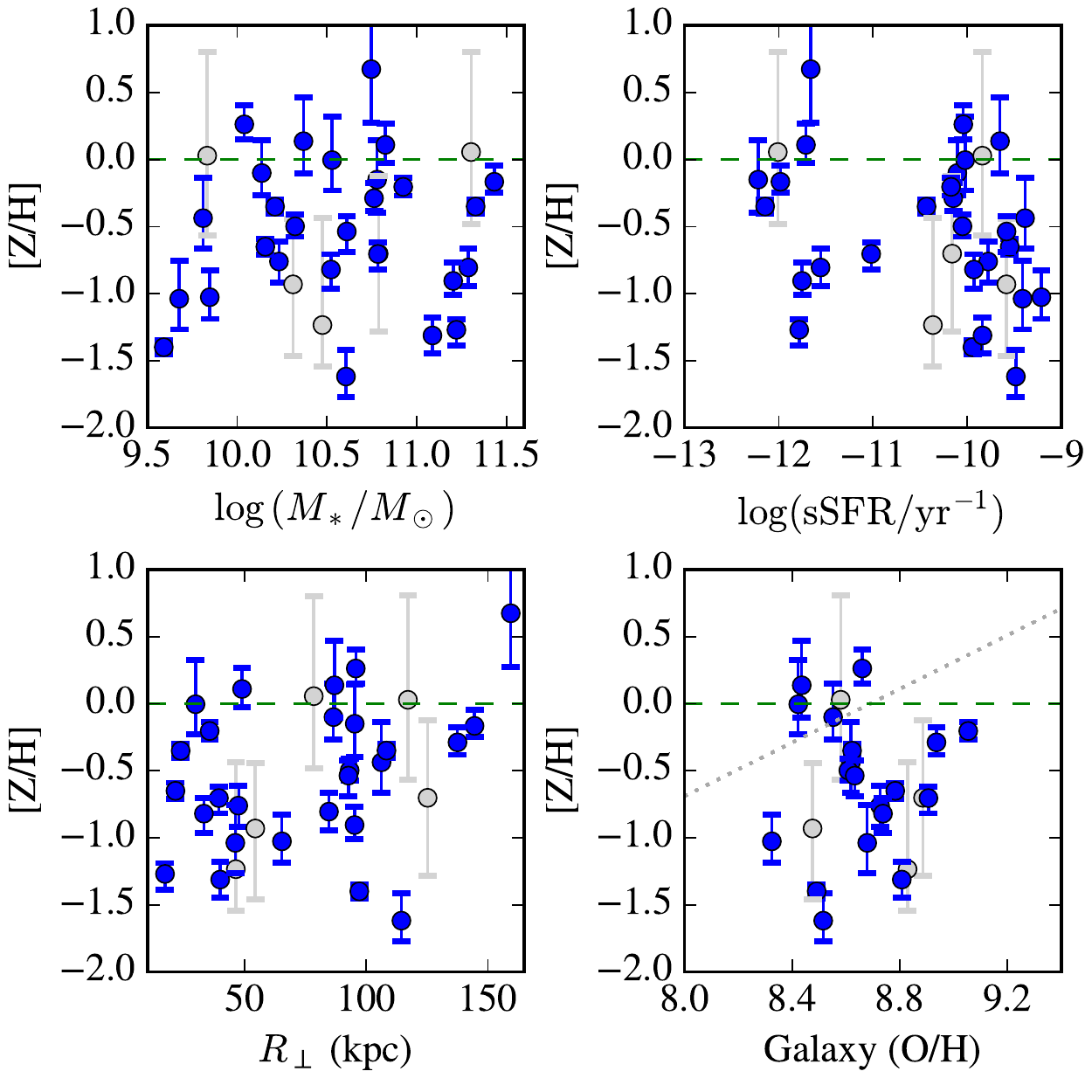}
\end{center}
\caption{
Comparison of the gas metallicity against several extrinsic
properties of the CGM systems.
There is little evidence for a correlation between [Z/H] and any
of these quantities, although there is a weak trend with \rperp.
This follows from the strong correlations between \nhi\ and \rperp\
and [Z/H] and \nhi. 
Grey points indicate systems with very poorly constrained values.
}
\label{fig:extrinsic}
\end{figure*}

At $\mnhi \approx 10^{15} \cm{-2}$, there are 
$\sim 3$ systems without a detection of Si$^{++}$
or C$^{++}$.  Most of 
these have $\N{Si^{++}} < 10^{12.5} \cm{-2}$
($2\sigma$), which is lower than the typical detection
but several have positive C$^{++}$ detections.
Furthermore, the addition of a few [Z/H]~$\lesssim -1$ systems
to Figure~\ref{fig:intrinsic} at low \nhi\ values would not 
qualitatively alter the observed trend.
We conclude that if one restricts to systems with 
$\mnhi > 10^{14.5} \cm{-2}$, then an anti-correlation 
exists between the enrichment level and the 
\ion{H}{1} column density in the  CGM of low $z$, massive galaxies,
under the assumption that photoionization equilibrium holds
over this range of \nhi.



\subsection{Extrinsic Trends}

In Figure~\ref{fig:extrinsic},
we examine trends of [Z/H] with a set of extrinsic
parameters.  
The stellar mass $M_*$, specific star formation
rate (sSFR), and nebular emission-line metallicity
measurements (O/H) are taken from \cite{werk12a}.
For the latter, we adopt their M91 calibration.

The [Z/H] vs.\ $M_*$ figure exhibits
no hint of an underlying trend.
There is, however, a weak, anti-correlation with the 
specific star formation rate (sSFR; null hypothesis
ruled out at $95\%$\ for the Pearson's test) and
a tentative positive correlation with
impact parameter ($96\%$).  
The latter follows
from two key results of this paper:
 (i) decreasing \nhi\ values with increasing impact parameters;
 (ii) an anti-correlation between \nhi\ and [Z/H].
The \rperp/[Z/H] correlation is at a lower statistical
significance, however, due to the large [Z/H] scatter
at all \rperp.
An anti-correlation between [Z/H] and sSFR
may run contrary to the interpretation
that the dependence of \ion{O}{6} on sSFR \citep{ttw+11}
is driven by metal-rich outflows \citep[e.g.][]{sbp+12}.

\subsection{Enhanced $\alpha$/Fe}
\label{sec:afe}

\cite{QPQ8} have recently reported enhanced ratios
$\alpha$-chain elements O, Si to Fe relative to the solar abundance
in the CGM surrounding massive galaxies at $z \sim 2$.
Their analysis is similar to the one presented here:  measurements
of ionic column densities (primarily low-ion transitions, e.g.\ \ion{O}{1}~1302,
\ion{Si}{2}~1304) converted to elemental abundances via corrections
from constrained photoionization models.  
Such an $\alpha$-enhancement may be expected for the gas surrounding
massive galaxies if the nucleosynthesis is dominated by Type~II supernovae.
We have compared our unenhanced models against the observed Si/Fe
ionic ratios and find no significant inconsistency.
At present, we find no evidence for an $\alpha$/Fe enhancement,
but caution that the uncertainties may exceed any expected enhancement.


\begin{figure}
\begin{center}
\includegraphics[width=3.5in]{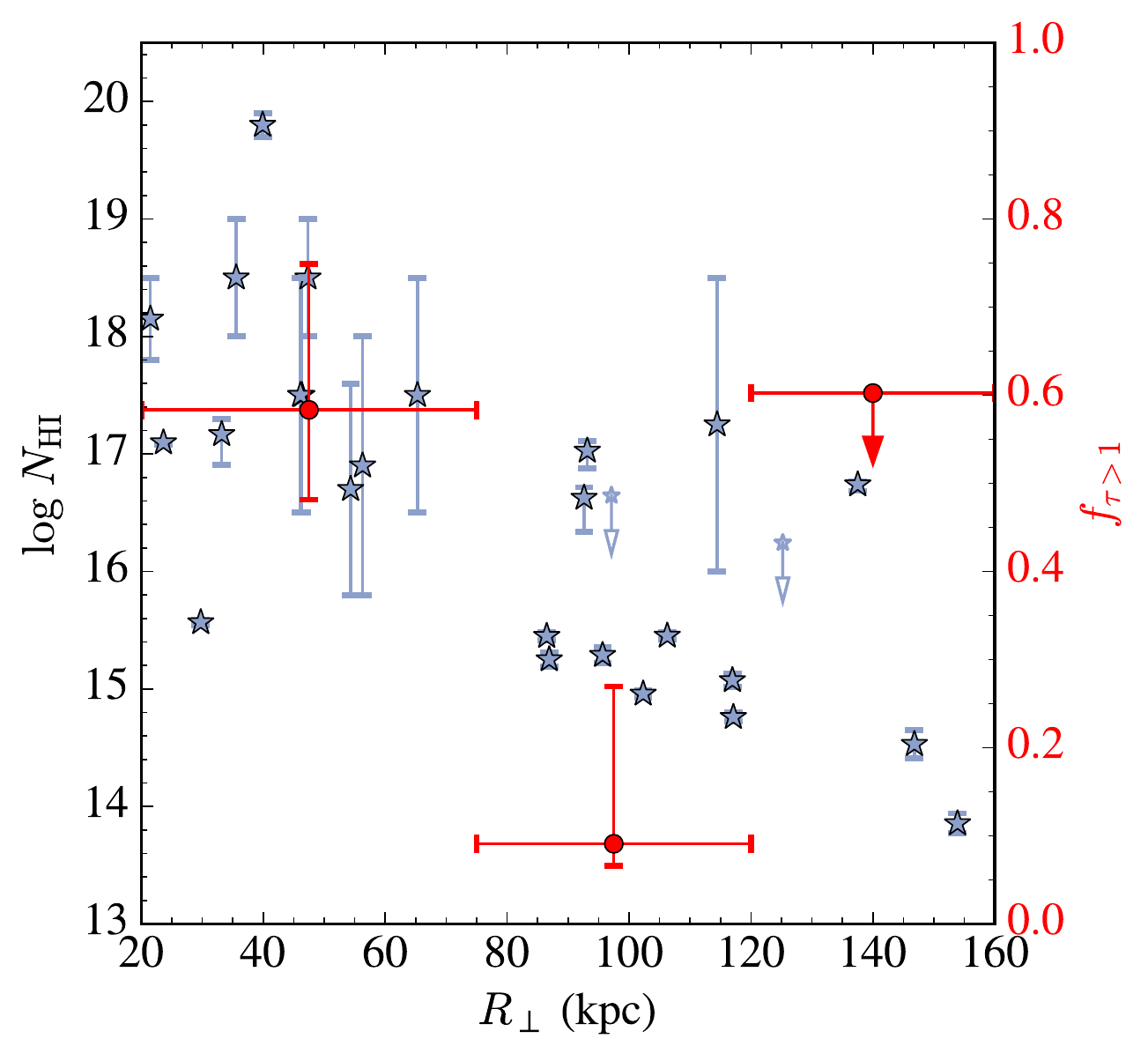}
\end{center}
\caption{
\nhi\ measurements vs. \rperp for star-forming galaxies within the COS-Halos survey. Arrows indicate upper limits. 
Overplotted in red are binned evaluations of \fecgm\ the fraction of systems 
with $\mtll \ge 1$. 
Uncertainties are standard binomial 68\%\ c.l. intervals while
the upper limit corresponds to 95\%\ c.l.
While the covering fraction of optically thick gas is high at 
$\mrperp \le 70$\,kpc, a significant fraction of sightlines has low opacity indicating that a non-negligible escape fraction through the CGM of 
$L^*$ galaxies.
}
\label{fig:fesc}
\end{figure}

\section{Discussion}
\label{sec:discuss}

We now discuss in greater detail the implications for 
several of the main results of this manuscript.
Throughout, we focus on the statistical ensemble
of COS-Halos measurements, and we remind the reader
that these are drawn from a homogeneous sample of 
sightlines penetrating the CGM of $z\sim 0.2$, field
$\sim L^*$ galaxies (i.e., $ 0.3 < L/L^*< 2$) with impact parameters $\mrperp < 160$\,kpc.

\subsection{Escape Fraction (\fecgm)}
\label{subsec:fesc}

Perhaps the dominant uncertainty in estimates of the 
$z<1$ EUVB is the contribution from star-forming galaxies
\citep[e.g.][]{hm01,kollmeier+14}.
This uncertainty stems primarily from the poor constraints
on the escape fraction $f_{\rm esc}$ of ionizing radiation
from the hot stars that produce these photons.
Most measurements have indicated a nearly negligible value
\cite[e.g.][]{leitherer+95}, but recent work has identified at least a subset of systems with significant leakage \citep{borthakur+14,izotov+16,leitherer+16}.

One of the contributing factors to the total \ftot\ value
is the CGM, i.e.\ the incidence of optically thick gas in
galaxy halos.  We may assess \fecgm, the escape fraction through the
CGM of star-forming $L^*$ galaxies at $z \sim 0$,
as follows.  Figure~\ref{fig:fesc} shows the \nhi\ measurements
versus \rperp\ for the star-forming galaxies of the COS-Halos
survey.  In three arbitrary \rperp\ bins we have calculated
\ftau, the fraction of sightlines
with $\mnhi \ge 10^{17.2} \cm{-2}$
corresponding to a Lyman continuum opacity $\mtedge \ge 1$.
The \ftau\ values and two-sided confidence intervals (68\%)
are overplotted on the data.  For $\mrperp < 75$\,kpc,
\ftau\ likely exceeds 0.5 however $\approx 40\%$ of the 
sightlines have $\mtedge \lesssim 1$.
This includes three sightlines with $\mrperp \lesssim 30$\,kpc,
implying that the CGM is not entirely opaque to ionizing
radiation.

We may estimate \fecgm\ from \ftau\ as follows.  First
we emphasize that a given CGM sightline from our experiment
travels through the entire halo at \rperp\ but 
does not sample radii $r < \mrperp$.
The this means that our dataset only constrains \fecgm\
for $r > 30$\,kpc.  And, in contrast to ionizing sources 
lyat the center of the halo (i.e.\ within the galaxy),  \ftau\ 
corresponds to approximately twice the opacity that a photon 
would encounter if emitted from the center.  
Because \ftau\ is large only in the inner bin, we base our
estimate on it alone.  Specifically, we approximate \fecgm\ as:

\begin{equation}
\mfecgm = 1 - \mftau(\mrperp < {\rm 70\, kpc})/2 
\approx \vfesc
\end{equation}

An estimate of \fecgm\ for the Milky Way has been performed
using surveys of the high velocity 
clouds \citep[HVCs][]{weiner+01,bland+01,foxetal06,wakker15}.
Their analysis indicates \fecgm\ of a few to several
tens percent which  is much smaller than our estimate.
This apparent discrepancy suggests that a significant fraction
of the opacity is due to gas with $r < 30$\,kpc, which is 
consistent with distance estimates for many HVCs
\citep[e.g.][]{thom08,wakker08}. However, we cannot know how typical the Milky Way is in this regard, or how this opacity varies with galaxy mass. As COS-Halos is not sensitive to $r < 30$ kpc, and the fraction of optically thick systems appears to increase rapidly down to and inside this radius, it remains possible that $L^*$ galaxies do have small CGM escape fractions.
In any case,  if the total escape fraction
is nearly 0, i.e. $\mftot \ll \mfecgm$,
then sources of opacity within the ISM or the first 30\,kpc of the CGM dominate.



\subsection{Enrichment of the cool CGM}
\label{sec:enrich}

Detections of strong metal lines in $\sim L^*$ galaxy halos demonstrate that the CGM is enriched in heavy elements
\citep{bergeron86,clw01}.
Thus far, however, a robust metallicity 
distribution function (MDF)
has been stymied by small sample sizes,
heterogeneous sample selection, large uncertainties in the hydrogen gas content, and ionization corrections.
The COS-Halos survey and the new \nhi\ and ionization
analyses presented here address these issues,
allowing a first estimate of the CGM-MDF.

\begin{figure}
\begin{center}
\includegraphics[width=3.5in]{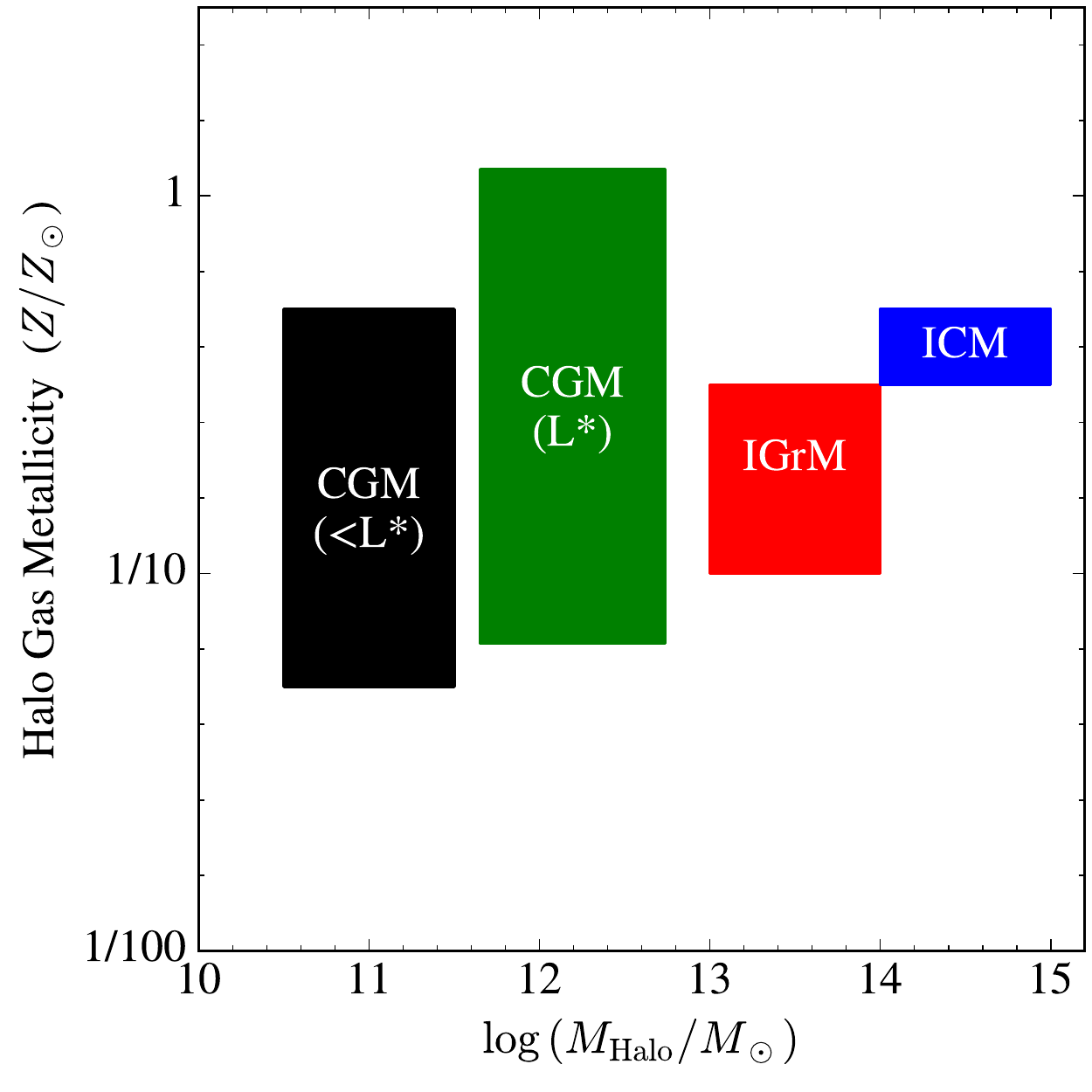}
\end{center}
\caption{
Halo gas metallicity plotted against total halo mass
for systems at $z \sim 0$.  The measurements were taken
from Bordoloi et al.\ (in prep; sub-$L^*$),
this work ($L^*$),
Rasmussen \& Ponman (2009; intragroup medium or IGrM),
and 
Maughan et al.\ (2008; ICM).
There is a general trend toward higher halo metallicity
with increasing mass although we stress that the individual
galaxies show a larger spread.
}
\label{fig:ZH_Mhalo}
\end{figure}

The primary result from the MDF  (Figure~\ref{fig:fullPDF}) is that the cool gas
within the CGM exhibits a metallicity exceeding 1/10 solar abundance. The median metallicity, measured from
the \nmetal\ COS-Halos systems analyzed, is
$\approx 1/3$ solar.  
This requires substantial and likely sustained enrichment
from the central galaxy and/or its progenitors.
This metallicity roughly matches
the values estimated for HVCs in our Galaxy
\citep[e.g.][]{gibson01,collins07} and new phenomenological models
for the hot halo \citep{faerman17}.

While the cases in which the CGM metallicity is higher than the metallicity derived from ionized gas within the galaxies
can potentially be understood by invoking metal-enriched outflows \citep{ps11},
the median CGM metallicity 
is significantly lower than the 
ISM metallicity
\citep[Figure~\ref{fig:extrinsic};][]{werk12a}.  
This indicates that the halo was 
primarily enriched by stars at an earlier time, when the galaxy itself had lower metallicity, or that metal-rich ejecta were diluted by more metal-poor 
gas within the halo, and/or lower metallicity gas from
accreting satellite dwarf galaxies \citep[e.g.][]{smg+13}.
We encourage the development of chemical evolution models
that focus on the CGM.

The median CGM metallicity is also 
consistent with the enrichment of the hot ($T > 10^6$K)
`halo' gas comprising the intracluster medium 
\citep[ICM; see Figure~\ref{fig:ZH_Mhalo};][]{maughan08}.
Indeed, the processes that polluted the CGM
of $L^*$ galaxies over the past $\sim 10$\,Gyr may be the same that enriched the ICM.  
In this picture,  the ICM represents the enriched halo gas stripped from $L^*$ galaxies and then shock-heated to the cluster virial temperature \citep[e.g.][]{mg95,siv09}.
In principle, this scenario could be tested by examining the
detailed abundance patterns of each.
Figure~\ref{fig:ZH_Mhalo} also shows current estimates
for the halo gas metallicity of the intragroup medium 
\citep[IGrM;][]{rp09} and estimates for the CGM of the sub-$L^*$
halos probed by the COS-Dwarfs survey (Bordoloi et al., in prep).
The IGrM and ICM suggest a trend toward higher metallicity at higher
halo mass.  The $L^*$ galaxies, however, exhibit a large spread
that extends even beyond the ICM measurements.  
Nevertheless, it appears reasonable that the halo gas of individual
galaxies can source the IGrM and ICM.

Our analysis detects no evidence for a radial gradient in the gas metallicity.  If anything,
[Z/H] {\it increases} at higher impact parameters 
(Figure~\ref{fig:extrinsic}).  
This may conflict with
models that envision the modern CGM to 
be dominated by on-going winds from the central galaxy. 
Instead, it may favor scenarios where the
CGM was polluted by one or more processes long ago
\citep{do07,ford+14,beno+16}\footnote{
One might also invoke enrichment by satellite galaxies, but we
note that \cite{burchett+16} found no excess of dwarf satellite galaxies near $z \sim 0$ \ion{C}{4} absorbers.
}.
Of course, this is most evident for the red-and-dead galaxies  of COS-Halos which also exhibit a high
metallicity CGM (log sSFR~$<-11$ in Figure~\ref{fig:extrinsic}). 

\begin{figure}[ht]
\begin{center}
\includegraphics[width=3.5in]{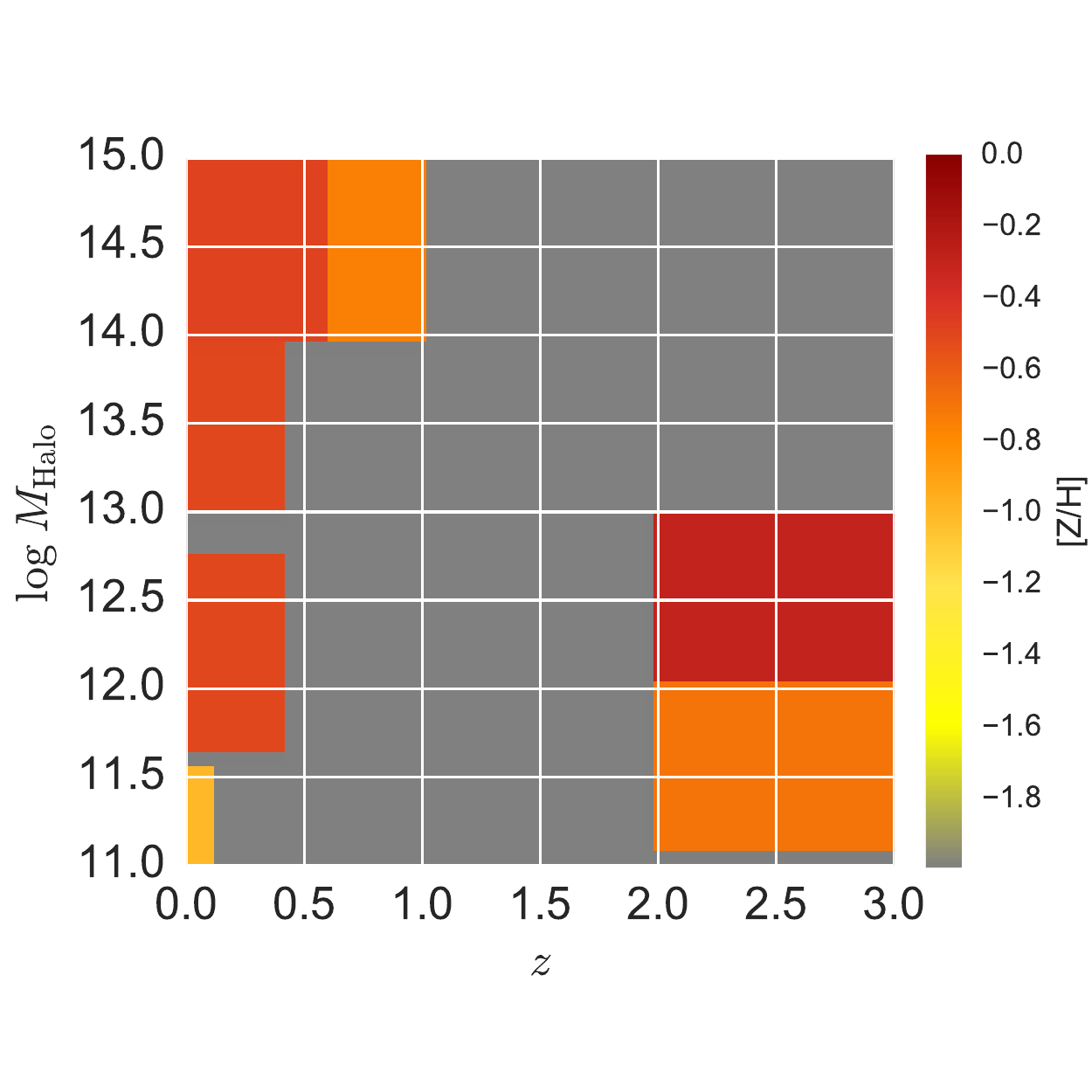}
\end{center}
\caption{
Graphic illustrating current constraints on the CGM enrichment
of dark matter halos as a function of mass and redshift.  This
includes our new results at low-$z$.  Ongoing surveys will
address the unconstrained areas (colored gray).
}
\label{fig:MzZ}
\end{figure}

We emphasize further that enriched gas is very likely to 
be present beyond the survey limit of 
COS-Halos (i.e. at $r>150$\,kpc)
in both the cool CGM \citep{zhu+14}
and the highly ionized gas probed by
\ion{O}{6} \citep{pwc+11,johnson+15}.
Such widespread and high metallicity implies an enrichment
process dominated by activity at early times.
One further appreciates that the CGM of bright $z \sim 2$ galaxies
also exhibits a high degree of enrichment
\citep[e.g.][]{crighton+15,turner+14,QPQ5,QPQ8}.
The terrific puzzle that emerges is whether we are observing
the {\it same} halo gas at $z \sim 0$ as observed at $z \sim 2$
\citep[see][for similar considerations 
but for \ion{O}{6} gas]{lehner+14}.  
Figure~\ref{fig:MzZ} expresses estimates for the metallicity
of the halo gas surrounding halos of a wide range of mass
and at varying redshift.

As is evident from Figure~\ref{fig:fullPDF} (and Figure~\ref{fig:ZH_Mhalo})
the CGM MDF for $L^*$ galaxies is broad, showing 
a 68\%\ c.l.\ interval of $\approx 1$\,dex.
Despite the large uncertainties to deriving metallicities from 
the (limited) observations of CGM systems, we contend that the
measured scatter includes a significant intrinsic contribution from metallicity variations within halos. 
This assertion is supported by Figure~\ref{fig:si3c3} where one
identifies large variations in $\N{Si^{++}}$ and $\N{C^{++}}$
at any given \nhi\ value.  Furthermore, we have argued for
examples of super-solar metallicity (Figure~\ref{fig:super})
yet expect these are a minority.
Unfortunately, we cannot yet test whether the dispersion is intrinsic
to individual halos \citep[see][for progress]{bowen+16}
-- thereby implying inefficient mixing 
\citep[e.g.][]{sck07} --
or tracks differences between halos.
On the latter point, we note no strong trends with
stellar mass (Figure~\ref{fig:extrinsic}) that could
generate an apparent dispersion.
Irrespective of its origin, the measured [Z/H]
dispersion places a new constraint 
on the physical processes that enrich the CGM.

\begin{figure*}[ht]
\begin{center}
\includegraphics[width=5.9in]{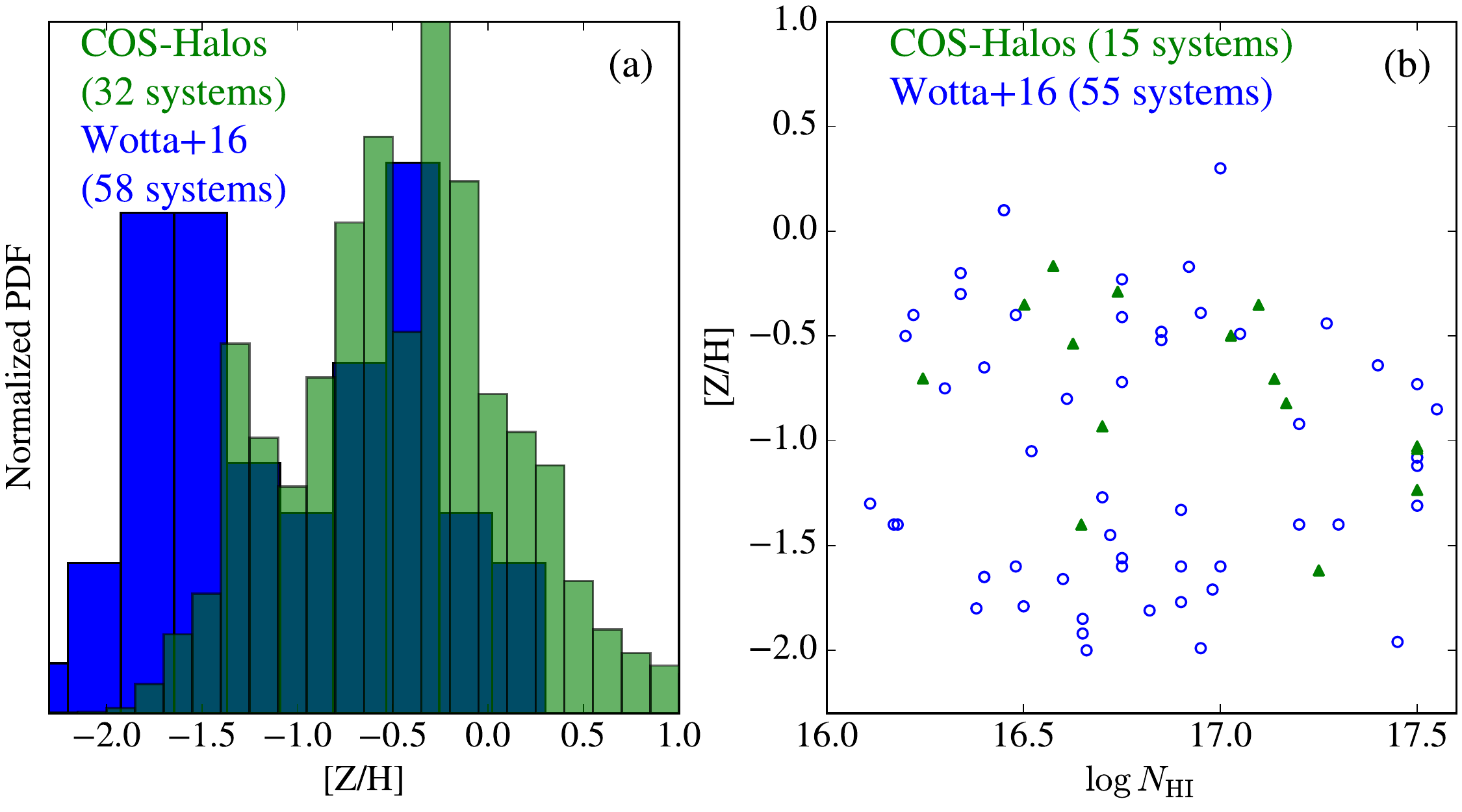}
\end{center}
\caption{
Comparison of the metallicity distribution functions (MDFs)
between the CGM of $L^*$ galaxies and the optically thick gas
traced by $z<1$ Lyman limit systems (LLSs; W16).
The left panel shows the complete samples (W16 limits are shown as values in this
presentation) while the right panel
is restricted to $\log \mnhi = [16, 17.6]$. 
In both panels we find that the $L^*$ CGM overlaps the high metallicity portion of the LLS MDF implying the former gives rise to the latter.  Furthermore, we propose that the lower metallicity LLSs might be associated with lower mass galaxies although no apparent trend with stellar
mass exists in our sample.
}
\label{fig:bimodal}
\end{figure*}

Lastly, we compare our results on the CGM
of $L^*$ galaxies with the 
MDF derived for $z<1$ Lyman limit systems\footnote{See
\cite{battisti+12} for higher \nhi\ systems.} 
\citep[LLSs;][]{lht+13,wotta+16},
which are also believed to trace the halos of galaxies
\citep[e.g.][]{lht+13,hafen+16}.
Figure~\ref{fig:bimodal} compares the MDF of the LLS analyzed
by  \cite[][hereafter W16]{wotta+16} 
against the full COS-Halos sample.
The COS-Halos MDF overlaps the higher metallicity 
measurements of the LLSs but shows a smaller
incidence of low metallicity gas.  
W16 have emphasized that the MDF of the LLSs is
bimodal when one restricts to the 
lower \nhi\ systems, aka partial LLSs or pLLSs.  
In the right panel of Figure~\ref{fig:bimodal}, we 
restrict both samples\footnote{Note that \cite{wotta+16} cut
their sample to focus on the partial LLSs, i.e., $\mnhi < 10^{17.2} \cm{-2}$.
The COS-Halos dataset has too few systems at those column densities to
enable a meaningful comparison, hence the higher \nhi\ cut here.}  
to $\mnhi = 10^{16}-10^{17.6} \cm{-2}$ and
see similar results to the full samples;
overlap at high [Z/H] and fewer CGM sightlines with
[Z/H]~$< -1$.  Performing a two-sided Kolmogorov-Smirnov
test on the sets of [Z/H] measurements rules out the
null hypothesis at $\approx 95\%$ that the two samples
are drawn from the same parent population.

We propose that a substantial 
fraction of the highly enriched, optically thick gas traced
by LLSs is associated
with $L^*$ galaxies.  Indeed, adopting the comoving number density
$n_{L^*}$
of $L > 0.5 L^*$ galaxies at $z \sim 0.3$ from \cite{loveday+15}
and $R_{\rm CGM} = 150$\,kpc,
and assuming the covering fraction of the CGM to pLLSs
to be $f_{\rm CGM}^{\rm pLLS} = 0.2$,
we predict an incidence: 

\begin{equation}
\ell(X) = n_{\rm L^*} \, \pi R_{\rm CGM}^2 \, f_{\rm CGM}^{\rm pLLS}
\approx 0.1 \;\; .
\label{eqn:lox}
\end{equation}
This is $\approx 50\%$ of the incidence of $\tau \ge 2$ LLSs 
($\mnhi \ge 10^{17.5} \cm{-2}$) 
estimated by \cite{ribaudo11}.   We conclude that the enriched halos of 
$L \approx L^*$ galaxies can explain the majority of high metallicity
LLSs observed by \cite{lht+13} and W16.
First results on associating the LLSs to galaxies support this
assertion \citep{lht+13}, but not without exception.

The other important conclusion from Figure~\ref{fig:bimodal}b
is that
the low metallicity pLLSs are unlikely to arise from the CGM 
of $L^*$ galaxies.  
There are, however, two caveats:
 (1) the gas could arise primarily at $\mrperp > 150$\,kpc, i.e.
 beyond the COS-Halos survey design \citep[although high \nhi\ values
 are more rarely observed at these separations][]{lht+13};
 and
 (2) the median redshift of the W16 sample 
 is $\tilde{z} \approx 0.6$,
 i.e. sampling an epoch 3.3\,Gyr earlier than the COS-Halos sample.
At a constant \nhi, one expects to probe higher overdensities
in our present-day universe.
Nevertheless, we suggest that the low metallicity gas observed by W16 is associated with the halos of lower mass galaxies
\citep[e.g.][]{rlh+11}, 
and further caution that it need 
not be linked to gas freshly accreting from the IGM. 


\begin{deluxetable}{ccc}
\tablewidth{0pc}
\tablecaption{CGM MASS\label{tab:mass}}
\tablehead{\colhead{$\mrperp^a$} & \colhead{$\log \mmcA$}
& \colhead{$\sigma(\log \mmcA)$}
\\ 
 (kpc) & ($M_\odot$) & ($M_\odot$)  } 
\startdata 
20 &8.9 &0.3\\ 
30 &9.9 &0.4\\ 
40 &9.4 &0.3\\ 
50 &8.2 &0.3\\ 
60 &10.0 &0.3\\ 
70 &7.3 &0.3\\ 
80 &10.6 &0.4\\ 
90 &9.2 &0.2\\ 
100 &8.2 &0.2\\ 
110 &10.2 &0.4\\ 
120 &9.0 &0.3\\ 
130 &8.6 &0.3\\ 
140 &8.1 &0.3\\ 
150 &8.4 &0.3\\ 
\hline 
\enddata 
\tablenotetext{a}{Inner radius of 10\,kpc annulus.} 
\end{deluxetable}

\subsection{Revisiting the Cool CGM Mass (\mcgm)}
\label{sec:mass}

The primary result of W14 was an estimate of the cool gas
mass of the CGM \citep[see also][]{stocke13}, as assessed
from a simple log-linear fit to estimates of $N_{\rm H}$
versus \rperp.  This analysis was subject to substantial 
uncertainty stemming from the large uncertainties on \nhi,
the systematic uncertainties of ionization modeling,
and the simplicity of this $N_{\rm H}(\mrperp)$ profile.
With our analysis, we have greatly improved the \nhi\ 
measurements and we provide a more robust
assessment of the error in photoionization modeling.
These may provide a more accurate and precise estimate
of \mcgm.  In addition, we introduce a new non-parametric
approach to the mass estimate. 

\begin{figure}
\begin{center}
\includegraphics[width=3.5in]{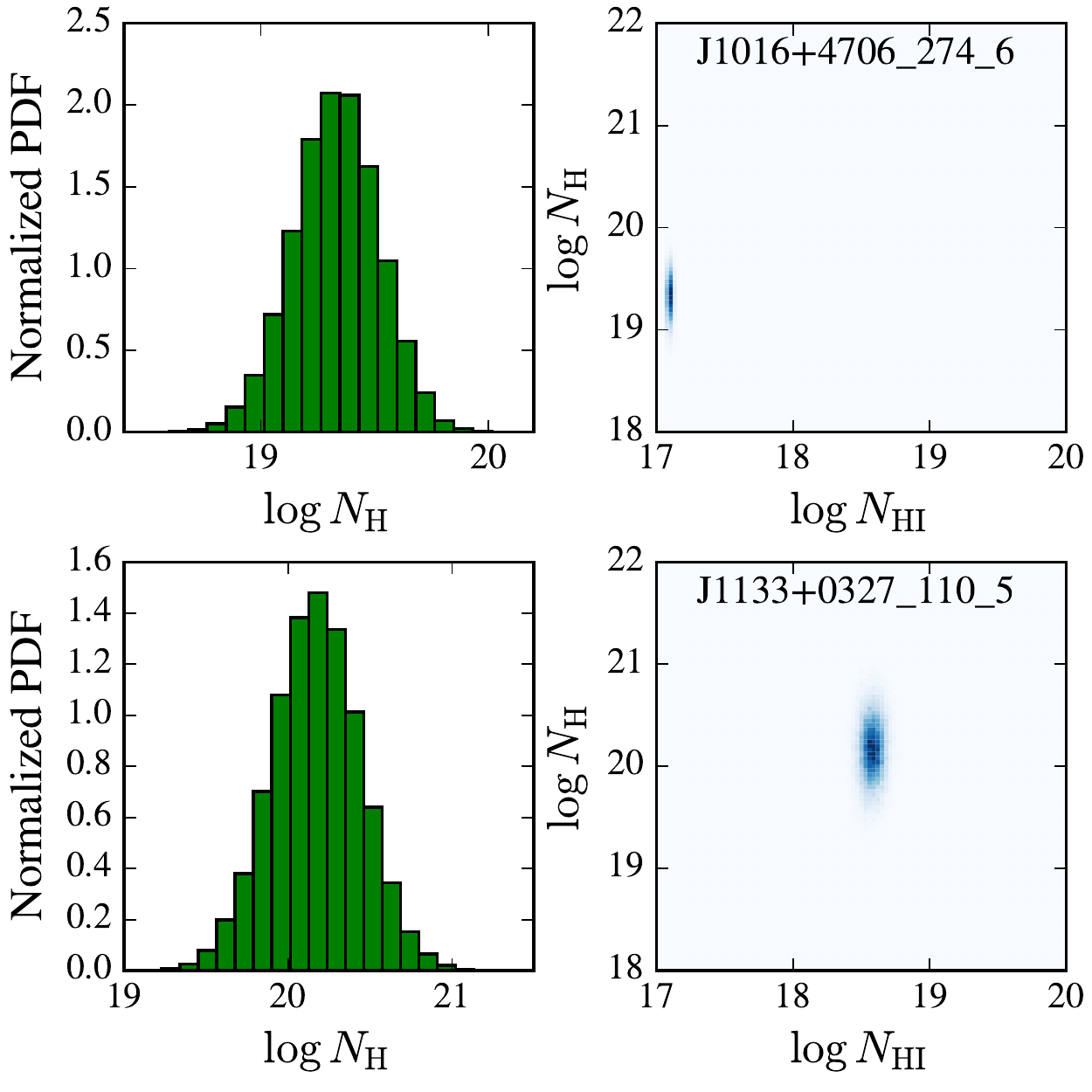}
\end{center}
\caption{
The left panels show the total hydrogen column density
\NH\ probability distribution functions (PDFs)
for two systems representative of the full sample.
These are derived from the MCMC analysis and also include
a 0.15\,dex Gaussian uncertainty from estimated systematic
error.  The right panels show the MCMC results in the
\NH/\nhi\ plane.
}
\label{fig:NH_exmp}
\end{figure}

Figure~\ref{fig:NH_exmp} shows the \NH\ PDF for two representative
systems, which differ greatly in the precision of their \nhi\ 
measurements.  The PDFs were generated from the MCMC ionization
analysis described in the Appendix 
and include an
additional \syserr\ Gaussian systematic uncertainty.
This systematic error dominates the PDF for J1016+4706\_274\_6
which otherwise exhibits a very narrow \NH\ distribution.
The uncertainty for J1133+0327\_110\_5, however, is 
dominated by the error in \nhi; one notes
a relatively tight correlation between the two properties.

\begin{figure}
\begin{center}
\includegraphics[width=3.5in]{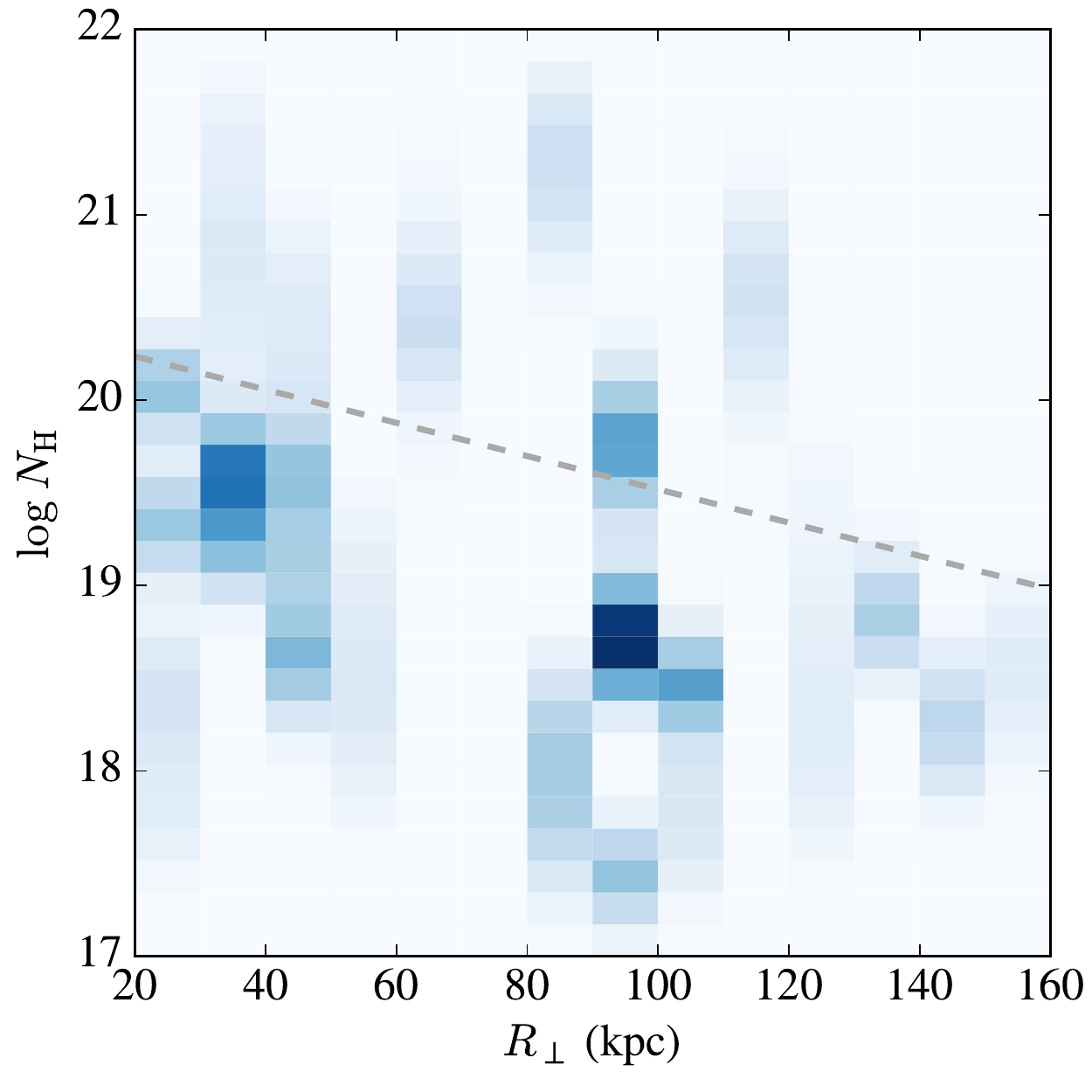}
\end{center}
\caption{
Two-dimensional histogram of the PDFs for the \nmetal\
systems of the COS-Halos survey analyzed here.
The bin sizes are 10\,kpc in \rperp\ and 0.17\,dex in \NH.
The darker colors reflect both the sharpness
of the PDF and the number of systems at a given \rperp\ bin.
The gray dashed line shows a fit to the \NH\ values versus
\rperp\ from W14.
}
\label{fig:NH_vs_R}
\end{figure}

By collating the \NH\ PDFs for the \nmetal~systems analyzed
from the COS-Halos survey, we may generate a 2D histogram in 
\NH-\rperp\ space (Figure~\ref{fig:NH_vs_R}).
Note that each system contributes equally to the histogram and that
several bins contain more than one system, i.e.\ the `maximum'
at $\mrperp \approx 90$\,kpc and $\mNH \approx 10^{18.8} \cm{-2}$
reflects both a sharply peaked PDF in that bin and the fact
that several systems contribute. 
A qualitative assessment of Figure~\ref{fig:NH_vs_R} suggests
a declining \NH\ value with increasing \rperp\ but also 
large \NH\ scatter both within and between the \rperp\ bins.
Future studies (e.g.\ the CGM$^2$ Gemini Large Program, PI Werk) should reduce the 
current sample variance.

We now offer a non-parametric estimate of the mass 
\mcgm\ of the cool CGM within 160\,kpc.
In \rperp\ bins
of $\Delta R = 10$\,kpc starting at 20\,kpc, we estimate a 
`best' \NH\ value \bNH\ and its uncertainty $\sigma(\mbNH)$.
Each bin then contributes an annular mass: 

\begin{equation}
\mmcA = m_p \, \mu\ \mbNH \, \pi [(R_{\perp,j} + \Delta R)^2 - R_{\perp,j}^2]
\cmma
\end{equation}
with $\mu \approx 1.3$ the reduced mass correcting
for Helium.  The 
total mass is trivially estimated by summing over
the annuli:

\begin{figure*}[ht]
\begin{center}
\includegraphics[width=5.9in]{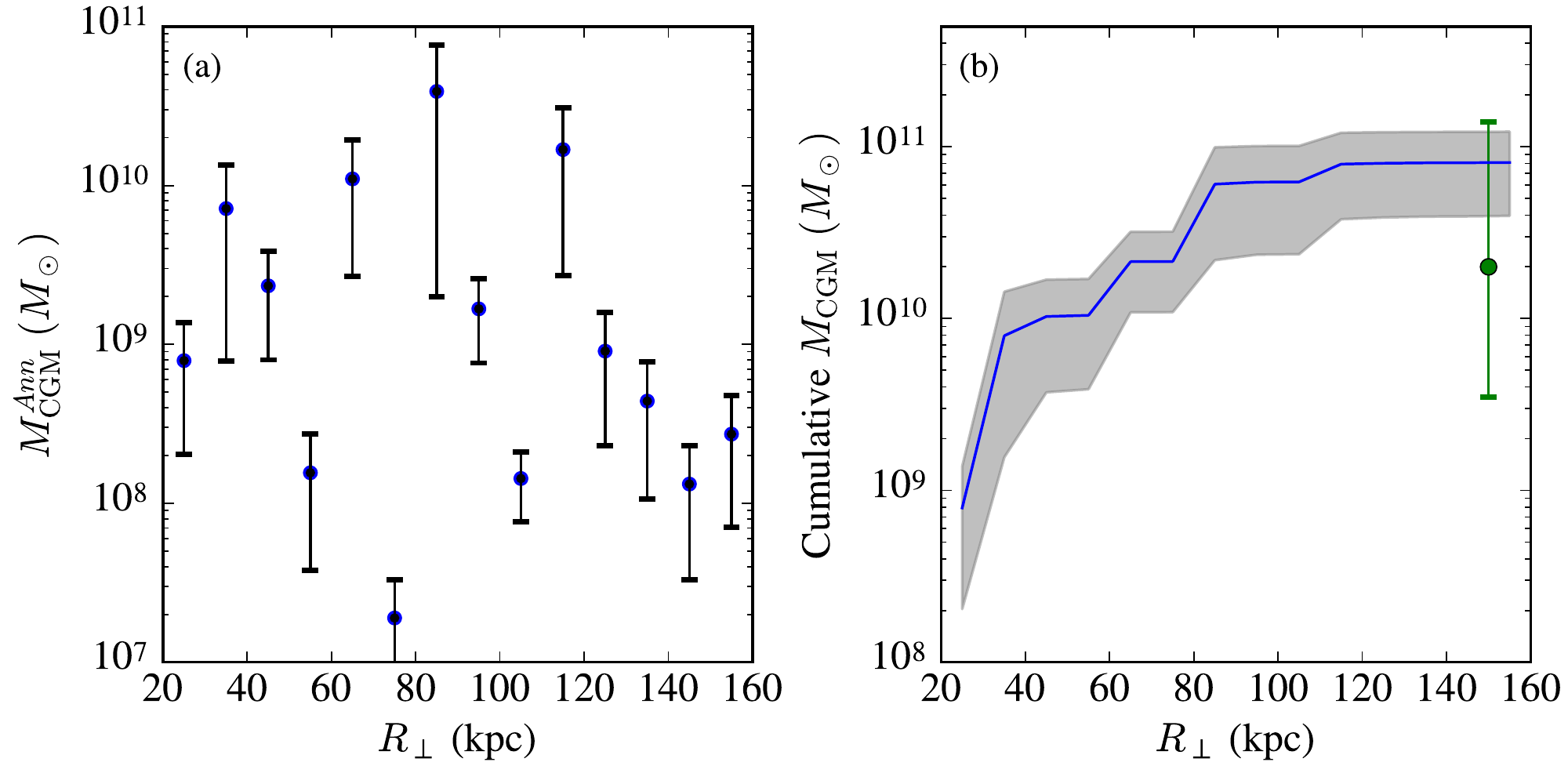}
\end{center}
\caption{
(a) Estimated mass in annuli of 10\,kpc width
for the CGM, estimated from the COS-Halos survey.
The uncertainty is estimated from a bootstrap
analysis (see text).  
(b) The estimated 
cumulative mass of the cool CGM gas. 
The green point with error bar shows the mass
estimate (with conservative bound) from W14 to
$\mrperp = 150$\,kpc.
}
\label{fig:MCGM}
\end{figure*}

\begin{equation}
\mmcgm = \smm_j \mmcA  \;\; .
\end{equation}
The challenge remains, however, to estimate \bNH\
and its uncertainty.  
There are at least three statistics one can derive
from a single \NH\ PDF: 
(1) the geometric mean $\mbNH = 10^{\langle \log \mNH \rangle}$;
(2) the true mean $\mbNH = \langle 10^{\log \mNH}\rangle$;
and
(3) the median. 
In practice, the first and last estimators yield similar
results because the PDFs are relatively symmetric in log space.
The true mean, however, yields systematically higher values
($\approx 0.3$\,dex).
Presently,  it is difficult to argue convincingly for any of these
prescriptions (on statistical or physical grounds), but consider
the following.   In the $\mrperp = [80,90]$\,kpc bin there 
are a pair of systems with \NH\ PDFs that peak at 
$\approx 10^{18}$ and $10^{21} \cm{-2}$.  Unless the high \NH\
system is a true statistical fluke, the average \NH\ value in
that annulus must be much closer to it.
Therefore, we proceed with the true mean and caution that 
the resultant mass estimate 
is especially sensitive to sample variance.


Figure~\ref{fig:MCGM}a shows the \mcA\ measurements vs.\
\rperp.  One finds a relatively flat profile 
which declines at higher \rperp\ values.  We have estimated
the uncertainty in each annulus by a two-fold bootstrap
procedure.  First, we randomly sample the \nmetal~systems
allowing for duplication.  Then we randomly sample each
system's \NH~PDF allowing for duplication.
We perform this exercise for 10,000 realizations and show
the standard deviation on the \mcA\ values
(Table~\ref{tab:mass}).  

Figure~\ref{fig:MCGM}b shows the cumulative mass profile.
Similarly, the uncertainty 
shows the standard deviation in the cumulative mass at
each \rperp\ bin.  
Altogether we estimate 
$\mmcgm = (\vmcgm \pm \emcgm) \sci{10} \msol$ 
to $\mrperp = 160$\,kpc.  Examining Figure~\ref{fig:MCGM},
it appears the mass has converged although this should be
confirmed by analyses at higher \rperp\ (e.g.\ extending
to the virial radius).  Our new estimate is consistent
with the lower limit established by W14.
It implies, as further emphasized in the next section,
that cool gas in the halo is a terrific reservoir of 
baryons, potentially rivaling the condensed baryonic
matter.

Lastly, we may perform the same analysis but weighting
$N_{\rm H}$ by the gas metallicity\footnote{In practice, we
draw from the [Z/H] values of the MCMC chains.  Also, we
assume an oxygen number abundance of 8.69 and that oxygen
represents 70\%\ of the mass in metals.}.  
This  provides an estimate of the metal mass in the
cool CGM,  $M_{\rm metals} = (\mmetal) \sci{8} \msol$.
This is higher than the estimates of W14 and
\cite{peeples+14} albeit with larger uncertainty.
Indeed, our central value even rivals the mass in
stars estimated by \cite{peeples+14}.
Further refining this mass estimate, therefore, 
bears directly on chemical evolution models for
galaxies like our own.

%
%
%

\subsection{Revisiting the Galactic Missing Baryons Problem}
\label{subsec:missing}

It has long been appreciated that the stars and ISM
of $L^*$ galaxies comprise far fewer baryons
\citep[e.g.][]{klypin+02}
than a simple scaling of the inferred total
halo mass \mhalo\ by the cosmic ratio of baryons to dark matter
$\rho_b/\rho_{\rm m} \approx 0.158$ \citep{wmap09}.
For a dark matter halo characteristic of the Milky Way
with $\mmhalo = 1.5 \sci{12} \msol$
\citep{bk13}, this implies a halo
baryonic mass of $\mmhalo^b \approx 2 \sci{11} \msol$. 
When first estimations of the mass of virialized gas
($T \gtrsim 10^6$\,K) suggested that 
$\mmhot \ll \mmhalo^b$  \citep[e.g.][]{anderson10},
researchers proposed that the halos hosting $L^*$ galaxies
were deficient in baryons yielding the so-called
galactic ``missing baryons problem''\footnote{This is frequently confused
with the intergalactic missing baryons problem \citep[see][]{fhp98}.}.
A more careful assessment of \mhot, however, showed that 
the uncertainties are large and systematically dependent on the
assumed mass profile \citep{fang+13} because
the most sensitive X-ray telescopes only probe the 
inner few tens kpc of distant galaxies.

Independent of the debate on \mhot,
estimates of the cool ($T \sim 10^4$\,K)
gas mass in the halo \mcgm\ derived from
CGM experiments indicate masses exceeding $10^{10} \msol$
\citep[][W14]{pwc+11,stocke13}.  In this manuscript, we have provided
a new estimate $\mmcgm = (\vmcgm \pm \emcgm) \times 10^{10} \msol$.
Obviously, this mass could resolve the galactic
missing baryons problem.  It would be astonishing and even
unsettling,  however, if $\mmcgm \gg \mmhot$.
At the same time, these same CGM experiments reveal a massive
reservoir of highly ionized gas traced by \ion{O}{6} absorption
\citep{pwc+11,ttw+11}.  Conservative estimates for the mass of the
highly ionized gas bearing O$^{+5}$ exceed $10^{9} \msol$,
assuming solar metallicity and physical conditions
that maximize the fraction of \ion{O}{6} \citep{ttw+11}. 
One then asks, how does \ion{O}{6}
relate to the hot halo, and is this highly ionized phase
a major baryonic component?  

One may gain special insight from observations of the Milky Way,
whose proximity affords a sensitive and unique perspective.
In particular, UV and X-ray observations provide absorption-line
measurements of the ionic column densities for 
O$^{+5}$, O$^{+6}$, and O$^{+7}$ along many sightlines to distant
sources \citep[e.g.][]{sws+06,fang+15}.
Furthermore, one observes the gas through X-ray
emission measurements \citep[e.g.][]{rp09}.
\cite{faerman17} have recently combined these constraints
to build a phenomenological model of the hot Milky Way halo
finding $\mmhot \approx 1.3 \sci{11} \msol$
\citep[see also][]{gupta+12}.  
This estimate is driven by two values: 
 (i) the characteristic column density of O$^{+6}$ which the
 community agrees is $N_{\rm OVII} \approx 2 \sci{16} \cm{-2}$,
 and
 (ii) an assumed spatial distribution $\ell_{\rm hot}$ for the hot gas.  
The former number is considered secure, and is only 1/2 the value one would
(presumably) measure along sightlines penetrating the entire halo.
The latter quantity, meanwhile, is hotly debated.

We emphasize first that the measured \ion{O}{7} column density
greatly exceeds the \ion{O}{6} measurements, 
i.e.\ $\N{O^{+6}}/\N{O^{+5}} \approx 100$.
Furthermore, there is evidence that the \ion{O}{6} gas is
distributed to hundreds of kpc ($\ell_{\rm OVI} > 100$\,kpc)
for our Galaxy \citep{sws+06,zheng+15} 
and external galaxies \citep{pwc+11,ttw+11,lehner+15}.
If the \ion{O}{7} gas is similarly distributed 
($\ell_{\rm OVII} \approx \ell_{\rm OVI} = \ell_{\rm hot}$),
a simple and large mass estimate follows:

\begin{equation}
\begin{split}
\mmhot \approx 10^{11} \msol \; \ltp \frac{f_{\rm OVII}}{0.5} \rtp^{-1}
\ltp \frac{\ell_{\rm OVII}}{100\,\rm kpc} \rtp^2 \\
\times 
\ltp \frac{\N{O^{+6}}}{4\sci{16} \cm{-2}} \rtp
\ltp \frac{Z}{0.5 Z_\odot} \rtp^{-1}
\end{split}
\end{equation}
where we assumed a correction for Helium and that
the logarithmic solar abundance of oxygen is 8.69, and
we adopted
conservative values for the \ion{O}{7} fraction $f_{\rm OVII}$
and the gas metallicity $Z$.
This estimate hinges on the value of $\ell_{\rm OVII}$
which \cite{faerman17} argue must be large to explain the 
observed X-ray emission.

On the other hand, \cite{yao07a} have interpreted the high
covering fraction of Galactic \ion{O}{7} absorption as evidence for a 
hot, thick disk with scale height of $\approx 1$\,kpc.
They found that they could reproduce the absorption and emission data
toward MRK~421 provided they also allowed for a non-isothermal
temperature profile.  They then argued that this disk
scenario should be favored over a Galactic halo origin for \ion{O}{7}
and \ion{O}{8} because
 (i) the halo gas should have low or even pristine metallicity;
 and
 (ii) the high incidence of \ion{O}{6} absorption toward
 distant sources favored a disk origin.  
We now appreciate, however, that the \ion{O}{6} gas is distributed on 
100\,kpc scales around galaxies 
\citep[including the Milky Way and Andromeda;][]{sws+06,zheng+15,lehner+15} 
and that the gas metallicity is far from pristine (e.g.\ Figure~\ref{fig:fullPDF}).
\cite{yao07a} further cited the lack of extended X-ray emission 
from the halos of external galaxies as evidence against that scenario,
but these measurements are not especially constraining.
At present, we find no reason to favor a disk origin for the hot
gas especially in light of the ubiquitous presence of \ion{O}{6} gas in
galaxy halos.

One path forward to assess $\ell_{\rm OVII}$ is to perform
a survey for strong \ion{O}{7} absorption  along quasar sightlines.  Following Equation~\ref{eqn:lox},
if $L^*$ galaxies exhibit strong \ion{O}{7} absorption to
$\ell_{\rm OVII}  = 100$\,kpc with a unit covering fraction,
then $dN/dz \approx 1$.
Unfortunately, the total redshift path surveyed to 
date is $\Delta z < 1$ \citep{fang06}
with one or two extragalactic \ion{O}{7} absorption systems 
detected \citep{nicastro+16}.  This supports scenarios
with a large $\ell_{\rm OVII}$, but any such conclusion is
tempered by sample variance.

An alternative and promising approach to statistically measure
the mass of ionized gas within galaxy halos is via the thermal
Sunyaev-Zeldovich effect \citep[SZ;][]{sz72}.
The most comprehensive measurement to date was performed by
the Planck Collaboration who examined 260,000 bright galaxies
associated with dark matter halos with $M_{\rm halo} > 2 \sci{13} \msol$
\citep{planck11}.  They report that a simple, single scaling relation
relates the SZ signal to galaxy mass down to stellar masses
$M_* \sim 2\sci{11} \msol$ and likely below  
\citep[see also][]{greco15}.
They further assert
that halos with masses from the largest clusters ($\approx 10^{15} \msol$)
to $\approx 10^{13} \msol$
(and likely below) have the mean cosmic fraction of baryons. 
It is highly suggestive, therefore, that the galactic missing baryons problem
exists only in-so-far that we have not yet identified the true
proportion of halo gas in cool, warm, and hot phases.
Developing such models while aiming to reproduce the 
primary CGM observations should be the focus of future work.

\section{Concluding Remarks}


In this manuscript and previous papers on the COS-Halos survey,
we have presented several surprising findings on the properties of
halo gas surrounding field $L*$ galaxies at $z \sim 0.2$.  
This includes high metal enrichment (including super-solar
metallicities) to beyond 100\,kpc, a cool gas mass
$\mmcgm \sim 10^{11} \msol$ that rivals any
other baryonic component in the halo, and an unexpected
anti-correlation between \nhi\ and metallicity.  

All of these
results depend on our treatment of the ionized gas measurements,
i.e.\ ionization corrections using relatively simple models.
In fact, no self-consistent and successful model for the halo
gas of any galaxy currently exists.
Therefore, we are compelled to conclude this manuscript with
several words of caution as regards CGM analysis and the results
that follow.

First and foremost, the standard photoionization models adopted
here and throughout the literature are known to fail when applied
to a wider set of ions, i.e.\ those with ionization potentials
IP~$> 25$\,eV \citep[e.g.][]{Werk:16,haislmaier+16}.  
This inconsistency may signal an inaccurate radiation
field \citep{cantalupo11}, a complex density structure
\citep{stern+16},  and/or additional ionization mechanisms.
For the primary results of this manuscript -- cool gas metallicity
and mass -- the implications are difficult to predict, but
we emphasize that a significant systematic uncertainty is lurking.

Second, we have yet to establish whether the lower ionization state
gas is in ionization or thermal equilibrium nor whether it is at
hydrostatic equilibrium within the underlying dark matter
gravitational potential.  We observe a wide range of
ionization states and infer multiple gas phases yet have not
developed even a simple model consisting of such phases
in pressure equilibrium.
Constructing such models for the CGM should be much easier
than theories of the ISM: one may largely ignore supernovae
energy/momentum input, molecules and dust are minimal,
star formation may be ignored, and magnetic fields may
play a small role.
Progress could and probably should follow a path similar to 
modeling of the ICM.

Lastly, we advise
observationalists (including ourselves)
to design experiments focusing on the astrophysics of the
medium.  Dedicated surveys with {\it HST}/COS and 10-m class
telescopes at $z \sim 2$ have yielded CGM datasets across 
cosmic time and for a diverse range of galaxies.  To faithfully
interpret these data, we must further constrain the underlying
astrophysics.  
This may be best achieved by accessing additional
absorption-line diagnostics \citep[e.g. OV, NeVIII;][]{tmp+11,mtw+13}
and higher spectral resolution
or by comparing the absorption-line data with extended
CGM emission.  And it may be as fruitful to return to 
our Galaxy and its nearest neighbors \citep[e.g. M31;][]{lehner+15}
where one can achieve exquisite sensitivity.

\acknowledgments
Support for program 13033 was provided by NASA through a grant from the Space Telescope Science Institute, which is operated by the Association of Universities for Research in Astronomy, Inc., under NASA contract NAS 5-26555. MSP and JT are partially supported by NSF grant AST-1517908 to the Johns Hopkins University. MF acknowledges support by the Science and Technology Facilities Council [grant number ST/L00075X/1]. N.T. acknowledges support from {\it CONICYT PAI/82140055}.
NL was partially funded by HST-AR-12854 from the Space Telescope Science Institute, which is operated by the Association of Universities for Research in Astronomy, Incorporated, 
under NASA contract NAS5-26555.
We thank  R. Dav\'e, Ben Oppenheimer, and Y. Faerman for helpful comments.
JXP thanks C. Wotta for including his dataset within pyigm.


\appendix

\section{Other \nhi\ Fits}

\begin{figure}
\includegraphics[width=6in]{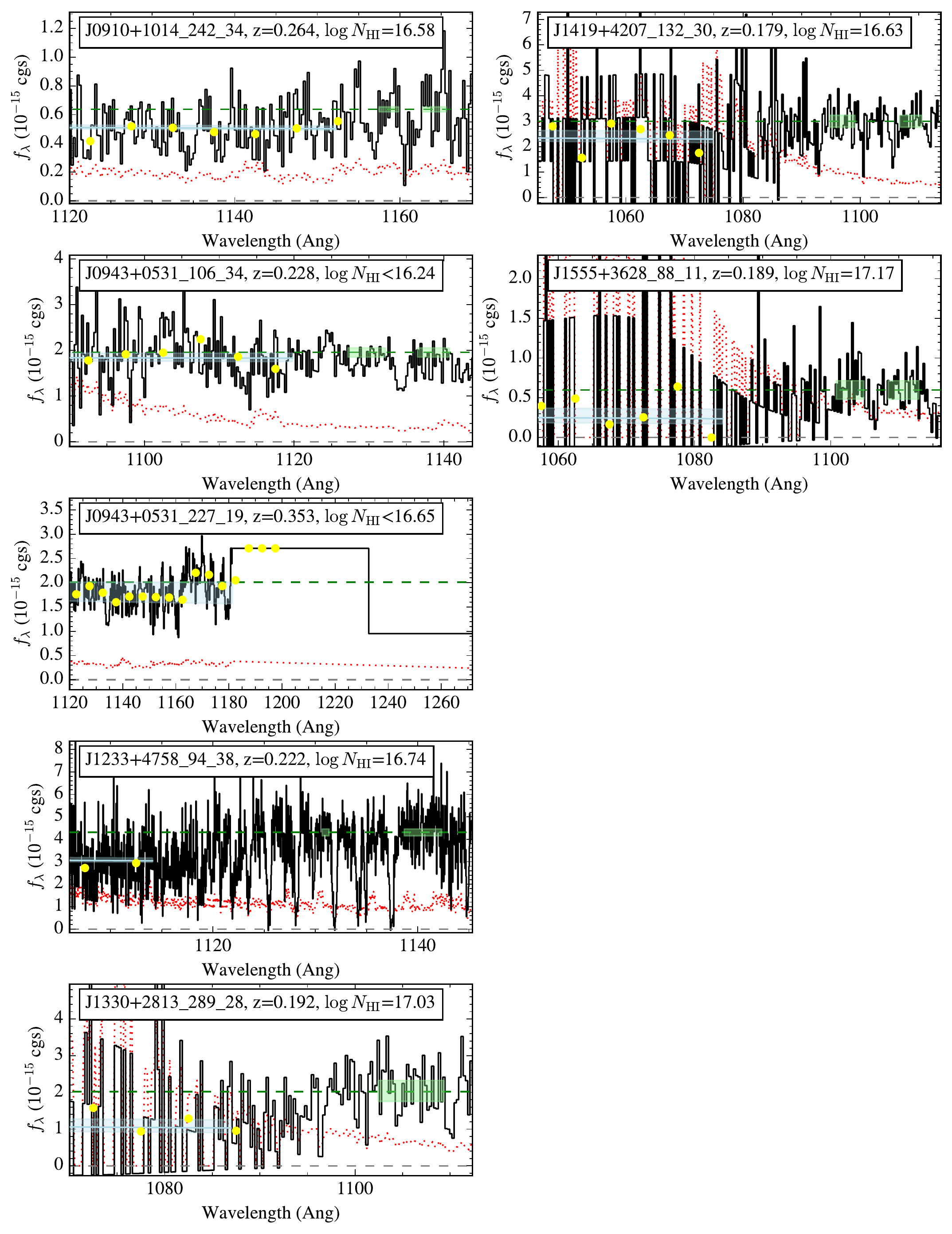}
\caption{
Same as Figure~\ref{fig:NHI_models} but for
the remainder of the sample.
}
\label{fig:other_NHI_models}
\end{figure}

The remainder of the systems analyzed at the
Lyman limit are presented in Figure~\ref{fig:other_NHI_models}.
The model parameters and fit results are given
in Table~\ref{tab:NHI_fits}.

\section{Ionization Modeling for Metallicity Evaluation}

In W14, we constructed photo-ionization models for \nwerk~sightlines in the COS-Halos sample.  Following standard practice, we compared the ionic column densities $N_{\rm ion}$ integrated over the full
system of low and intermediate ionization states 
(e.g.,\ Si$^+$, Si$^{++}$, Mg$^+$) against a grid of photoionization
models generated with the Cloudy software package \citep{ferland13}.  
Throughout the W14 analysis, we assumed the \cite{hm01} extragalactic
UV background (EUVB; HM2001)
radiation field and imposed the arbitrary
prior that the gas metallicity could not exceed the solar value,
which has been violated in several absorption systems in other studies \citep[e.g.][]{tmp+11,mtw+13}.
Constraints on the ionization model, specifically the ionization parameter $U$,
were assessed primarily through a visual comparison of the data to
models. Conservative estimates on the error in $U$ were adopted to account for this `by-eye' procedure and the simplifying assumptions
inherent to the photoionization modeling (e.g.,\ a 
constant density gas).

There are several differences between this analysis and 
W14. First, we have reassessed the measurements of ions in the COS spectra and redefined previously reported detections
as upper limits or as non-constraining due to unidentified blends
or poor data quality. Table~\ref{tab:updates} summarizes the modifications\footnote{The  entire COS-Halos database is now available as a tarball of JSON
files within the pyigm repository: https://github.com/pyigm/pyigm.
Software is included for ingesting these
data and performing meta-analysis. All of the spectra are bundled in v02 of {\it igmspec}, available for download
at https://specdb.ucolick.org
}. Second, we have ignored \ion{Mg}{1} throughout the analysis.
We have found that $\N{Mg^0}$ rarely offers a meaningful constraint and in a few cases yields highly conflicting results (especially in
systems with large \nhi).  Evidently our ionization models do
not capture an aspect of the astrophysics (e.g.,\ dust extinction, 
an unresolved colder phase)
or atomic physics (e.g.\ recombination coefficients) relevant
to \ion{Mg}{1}. Third, we have modeled our spectra using the most recent EUVB from \cite[][ HM2012]{hm12}, which exhibits a shallower slope than HM2001 between 1.5 $-$ 4 Ryd. In other words, HM2001 somewhat under-produces species with ionization potential energies between 1.5 and 4 Ryd  (e.g. SiIII) relative to the lower ionization potential ions (e.g. MgII, SiII) compared to HM2012. Overall, the difference is such that the gas ionization parameters derived from HM2001 will be $\sim$0.3 dex higher than those derived from HM2012 for the same sightlines. 

\begin{deluxetable}{lccccccccc}
\tablewidth{0pc}
\tablecaption{COLUMN DENSITY UPDATES\label{tab:updates}}
\tablehead{\colhead{System} & \colhead{Ion/Trans$^a$}
& \colhead{$f_{\rm orig}^b$} & \colhead{$f_{\rm new}^b$}
} 
\startdata 
J0910+1014\_34\_46& ${\rm N}^{+}$ & 1 & 3\\ 
J0928+6025\_110\_35& FeIII 1122 & 2 & 1\\ 
J0943+0531\_227\_19& ${\rm N}^{+}$ & 2 & 3\\ 
& ${\rm C}^{+}$ & 1 & 3\\ 
J1016+4706\_274\_6& FeII 1144 & 1 & 3\\ 
J1342-0053\_157\_10& OI 971 & 1 & 3\\ 
J1435+3604\_68\_12& OI 971 & 1 & 3\\ 
J1619+3342\_113\_40& ${\rm C}^{+}$ & 1 & 3\\ 
J2345-0059\_356\_12& NII 1083 & 1 & 3\\ 
& SiIII 1206 & 1 & 0\\ 
\hline 
\enddata 
\tablenotetext{a}{Original flag on the measurement (0=Not included; 1=Good measurement; 2=Lower limit; 3=Upper limit).} 
\tablenotetext{b}{Updated flag on the measurement.} 
\end{deluxetable}


Fourth, and most importantly, we adopt a Monte Carlo Markov Chain (MCMC) approach to compare an interpolated photoionization grid to the observational constraints from each system.  Full details of the procedure are provided in \cite{fumagalli+16a} and the code is publicly available\footnote{https://github.com/pyigm/pyigm}
and makes use of the {\tt EMCEE} package
\citep{emcee+13}. 
Here we briefly summarize the algorithm. We first generated a grid of equilibrium photoionization models (recovering $T \sim 10^4$\,K),
each with a constant gas density $n_{\rm H}$.  The gas
has solar relative abundances  \citep{asplund09},
scaled to a global metallicity [Z/H]. The grid has two additional parameters: the integrated \ion{H}{1} column
density \nhi\ and the redshift $z$.  The latter
sets the adopted radiation field which is taken to be the
extragalactic UV background (EUVB) derived from the CUBA 
package \citep{hm12}. The uncertainty in the EUVB intensity remains large \citep[e.g.][]{kollmeier+14} and this primarily affects
our density estimations. Systematic uncertainty in the
shape of the EUVB imposes a systematic error in the metallicity
of $\approx 0.3$\,dex \citep{howk+09,fumagalli+16a,wotta+16}.
The \nhi\ value sets the thickness of the plane parallel gas layers for each solution 
The ranges for the four grid parameters are summarized in Table~\ref{tab:param}.
For the two systems analyzed with $\log \mnhi < 15$, we 
ran the analysis assuming $\log \mnhi = 15.5$ and
afterwards offset accordingly the outputs.  At these low
\nhi\ values where the gas is optically thin to
ionizing radiation, the relative populations of the 
ionization states have very little \nhi\ dependence.

\clearpage
\begin{deluxetable}{lccccccccccccc}
\tablewidth{0pc}
\tablecaption{Cloudy Model Parameters\label{tab:param}}
\tablehead{\colhead{Parameter} & \colhead{Range} & \colhead{Step Size}}
\startdata 
[Z/H]                     & -4, 2.5  & 0.25 \\
z                         & 0, 4.5   & 0.25 \\
$\log \mnhi/\cm{-2}$      & 15, 20.5 & 0.25 \\
$\log  n_{\rm H}/\cm{-3}$ & -4.5, 0  & 0.25 \\
\hline
\enddata 
\end{deluxetable}

We emphasize that the models assume an overly simplified constant density for all gas layers. Recent work has demonstrated that relaxing this assumption may describe a wider range of the observed
ions with even fewer parameters \citep{stern+16}. On the other hand, we are strongly motivated to these `single phase' models by the tight kinematic correspondence between the \ion{H}{1} Lyman series and the lower ionization state gas \citep{Werk:16} and because these models
provide a good fit to the lower ionization state gas 
in the majority of cases \citep[see also][]{haislmaier+16}.

For each CGM system, we performed an initial run with the
MCMC randomly seeding the initial values
for the walkers throughout the full grid of model parameter space. We generated 960 walkers with 480 samples per MCMC chain (eventually removing a `burn-in' set of 45 samples per chain). For those systems with at least one measurement of an ion column density, 
the acceptance rate was approximately a nominal level of 0.5. 
The 9 systems without a metal constraint yielded a zero acceptance rate and are considered no further.


We then performed a second MCMC run seeded by the initial results.  
We initialized these chains at
the median values of the initial runs with a normal
deviate in log10 space of 0.01\,dex.
From this second run, we derive the final adopted probability distribution functions (PDFs) for the model parameters. 

\begin{figure}
\includegraphics[width=7in]{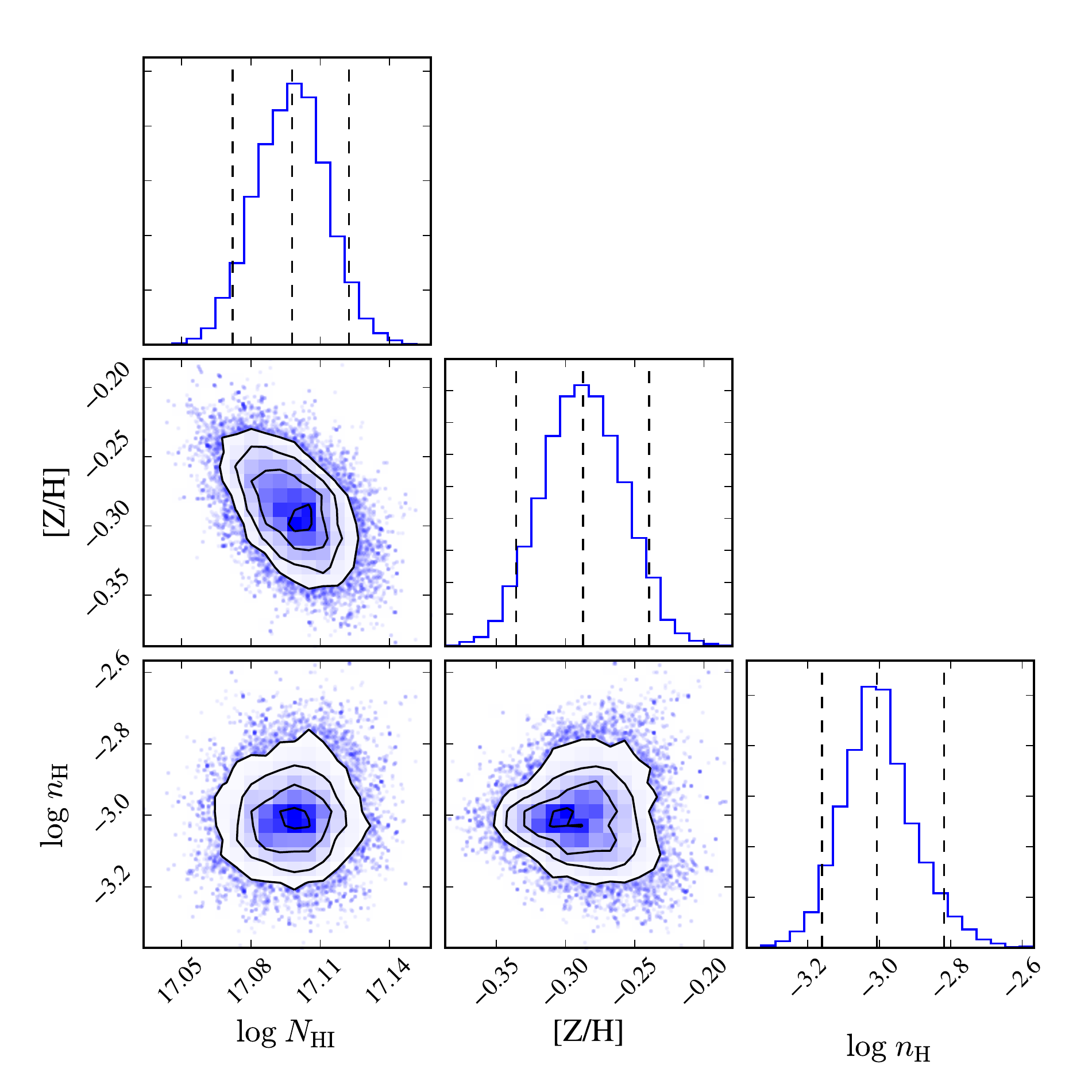}
\caption{
A `corner' plot of the MCMC PDF for the model parameters of 
J1016+4706\_274\_6.   This describes the distribution
of model parameters in the MCMC chains.
}
\label{fig:corner}
\end{figure}

\begin{figure}
\includegraphics[width=3.5in]{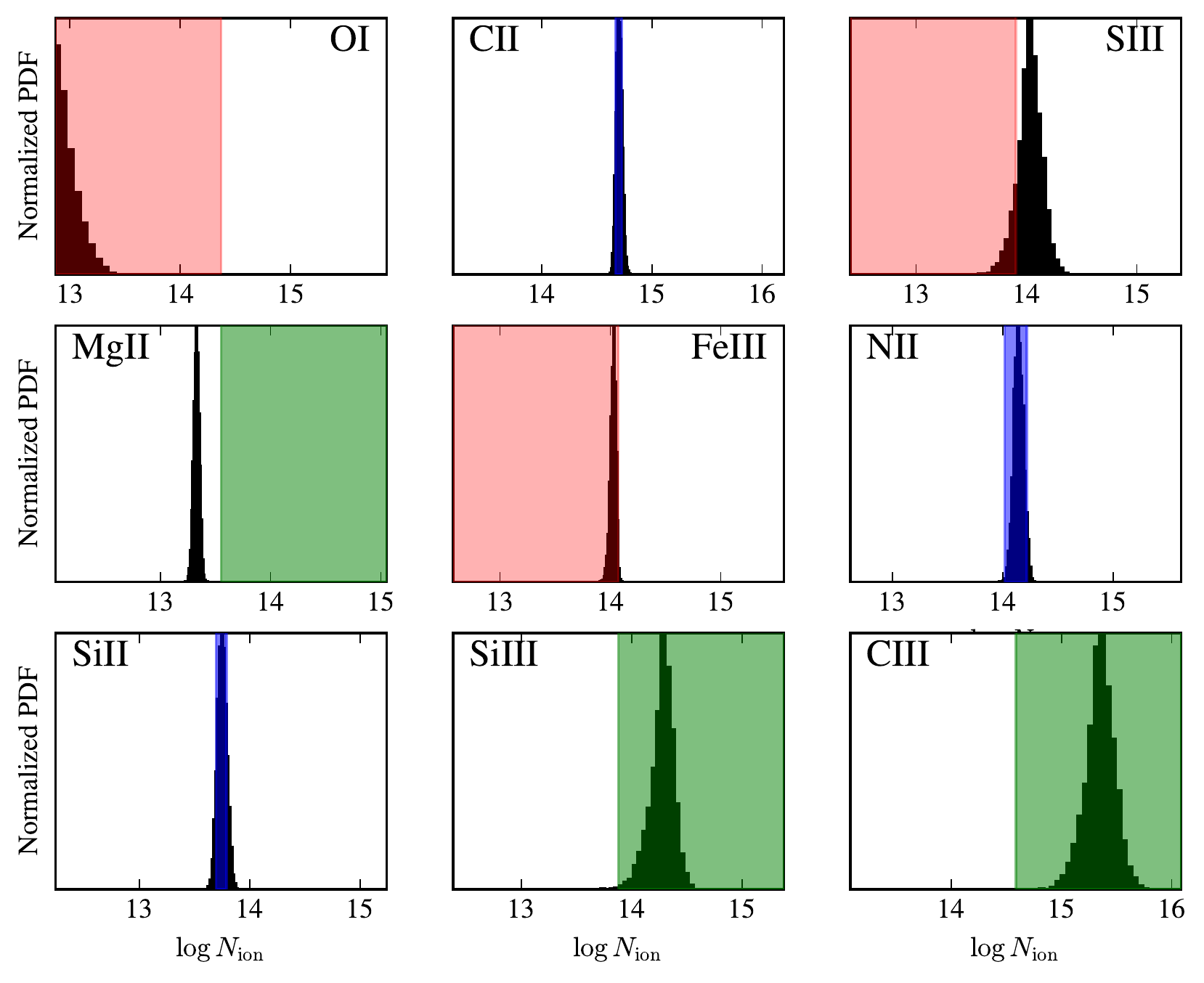}
\caption{
Each panel compares the model predictions 
for the ionic column densities $N_{\rm ion}$ against the observational constraints (ordered by ionization potential).  
Blue shaded regions
indicate a measured value for $N_{\rm ion}$ with $1\sigma$
uncertainty whereas the pink/green regions indicate
upper/lower limits on $N_{\rm ion}$ respectively.  
With the exception of \ion{Mg}{2}, where the model 
under-predicts the observed constraint, 
there is good agreement.
}
\label{fig:residuals}
\end{figure}

Figure~\ref{fig:corner} shows a corner plot for three
of the model parameters 
for a well-constrained system (J1016+4706\_274\_6).
We designate the preferred or `best' model from the
median of the parameter PDFs when discussing individual systems
and the uncertainties are based on percentiles of the PDF.
These quantities are well-behaved for this model.
Figure~\ref{fig:residuals} compares the
observational constraints with the model PDFs for the ionic
column densities.  All of the observables are well-modeled
with a slight tension for \ion{S}{3} and the
under-prediction of \ion{Mg}{2}. 
Such deviations from these species are common in absorption-line modeling
\citep[e.g.][]{pro99,crighton+15,
haislmaier+16,wotta+16},
and they suggest either over-simplifications in the modeling (e.g.\ 
constant density), non-solar relative abundances within the gas from
nucleosynthesis, and/or differential dust depletion.
For completeness, Table~\ref{tab:columns} provides the
measurements of each ionic column density used in the
analysis and the model results.

\begin{deluxetable*}{lccccccccc}
\tablewidth{0pc}
\tablecaption{IONIC COLUMN DENSITIES AND MODEL VALUES\label{tab:columns}}
\tabletypesize{\scriptsize}
\tablehead{\colhead{Galaxy} & \colhead{Ion}
& \colhead{$\log N$} & \colhead{$\sigma(\log N)^a$}
& \colhead{Model$^b$}
} 
\startdata 
J0401-0540\_67\_24& OI& 14.15& 99& 9.77,11.92\\ 
& SiII& 12.47& 99& 11.56,13.14\\ 
& CII& 13.58& 99& 12.99,13.77\\ 
& MgII& 12.26& 99& 10.87,12.39\\ 
& NII& 13.55& 99& 12.16,13.12\\ 
& FeII& 13.89& 99& 9.24,11.77\\ 
& FeIII& 13.85& 99& 11.63,12.79\\ 
& SiIII& 12.88& 0.06& 12.77,13.00\\ 
& CIII& 14.00& -1& 14.01,14.94\\ 
J0803+4332\_306\_20& OI& 14.17& 99& 7.21,14.86\\ 
& SiII& 12.68& 99& 9.58,13.82\\ 
& CII& 13.58& 99& 11.89,14.65\\ 
& MgII& 12.00& 99& 7.99,13.86\\ 
& NII& 13.68& 99& 10.79,13.86\\ 
& FeII& 13.58& 99& 6.36,13.79\\ 
& FeIII& 14.16& 99& 9.32,12.70\\ 
& SiIII& 12.48& 99& 10.86,12.66\\ 
& CIII& 13.67& 0.06& 13.54,13.78\\ 
\hline 
\enddata 
\tablenotetext{a}{Error in the column density measurement.  A value of -1 indicates a lower limit.  A value of 99 indicates an upper limit.}
\tablenotetext{b}{Range of model values the column density  (95$\%$ interval)}
\end{deluxetable*}

For comparison with other results from photoionization
modeling of absorption systems, we estimate
$\log U \approx -3.1$ for $\log n_{\rm H} = -3$ at
$z=0.2$ for our adopted EUVB where
$U \equiv \Phi/(c n_{\rm H}$ with $\Phi$
the flux of ionizing photons.
If one were to increase/decrease the intensity, e.g.\ a
local enhancement related to star-formation within
the galaxy, the first-order effect is a corresponding
increase/decrease in $n_{\rm H}$ because the relative
ionic column densities are most sensitive to $U$.

\begin{figure}
\includegraphics[width=3.5in]{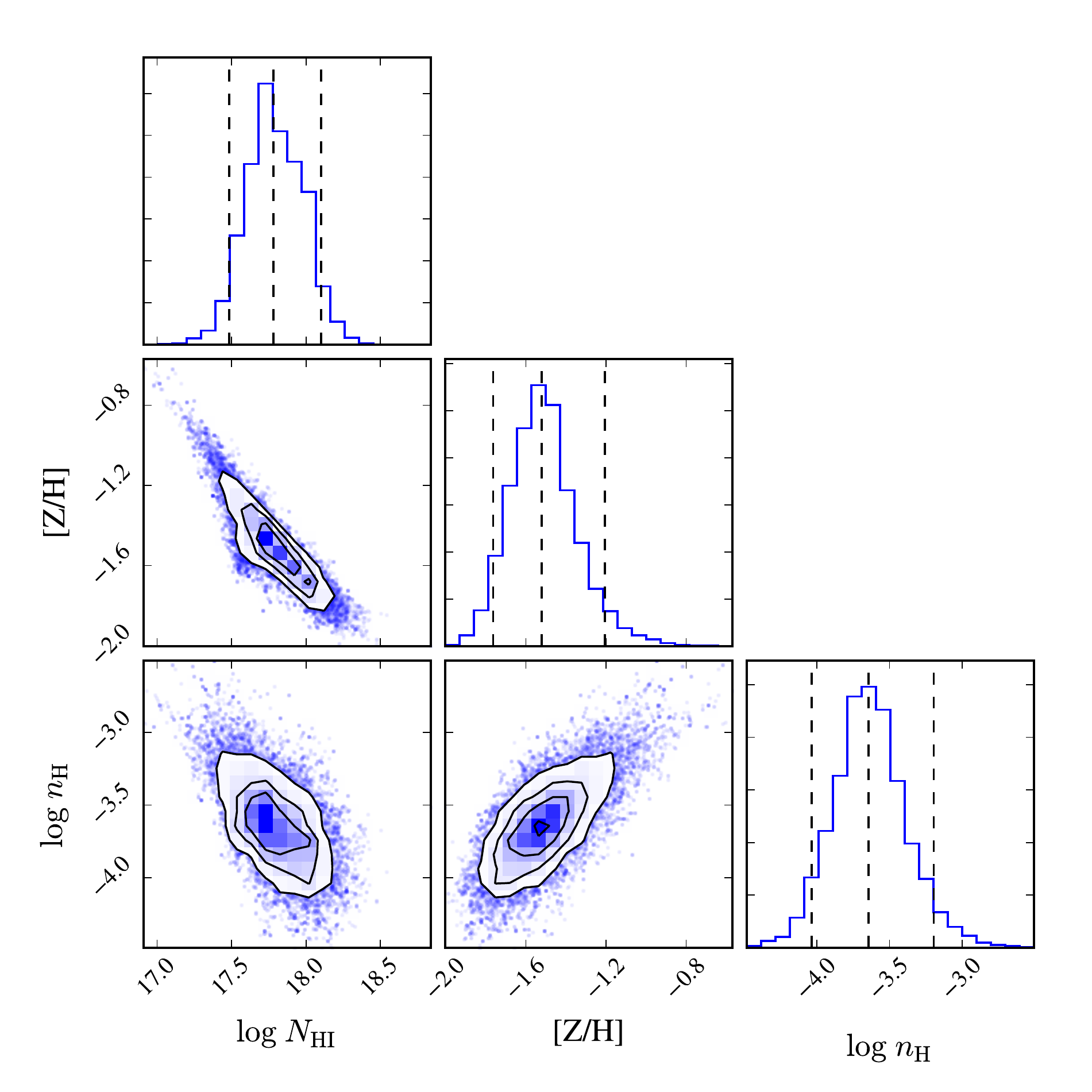}
\caption{
Same as Figure~\ref{fig:corner} but for a system with
fewer observational constraints. 
}
\label{fig:corner_appx}
\end{figure}

Figure~\ref{fig:corner_appx} shows another corner
plot for one of the MCMC models.  In contrast to
Figure~\ref{fig:corner}, this model has fewer
observational detections and the resultant
constraints on the model are poorer.

\begin{figure}
\begin{center}
\includegraphics[width=3.0in]{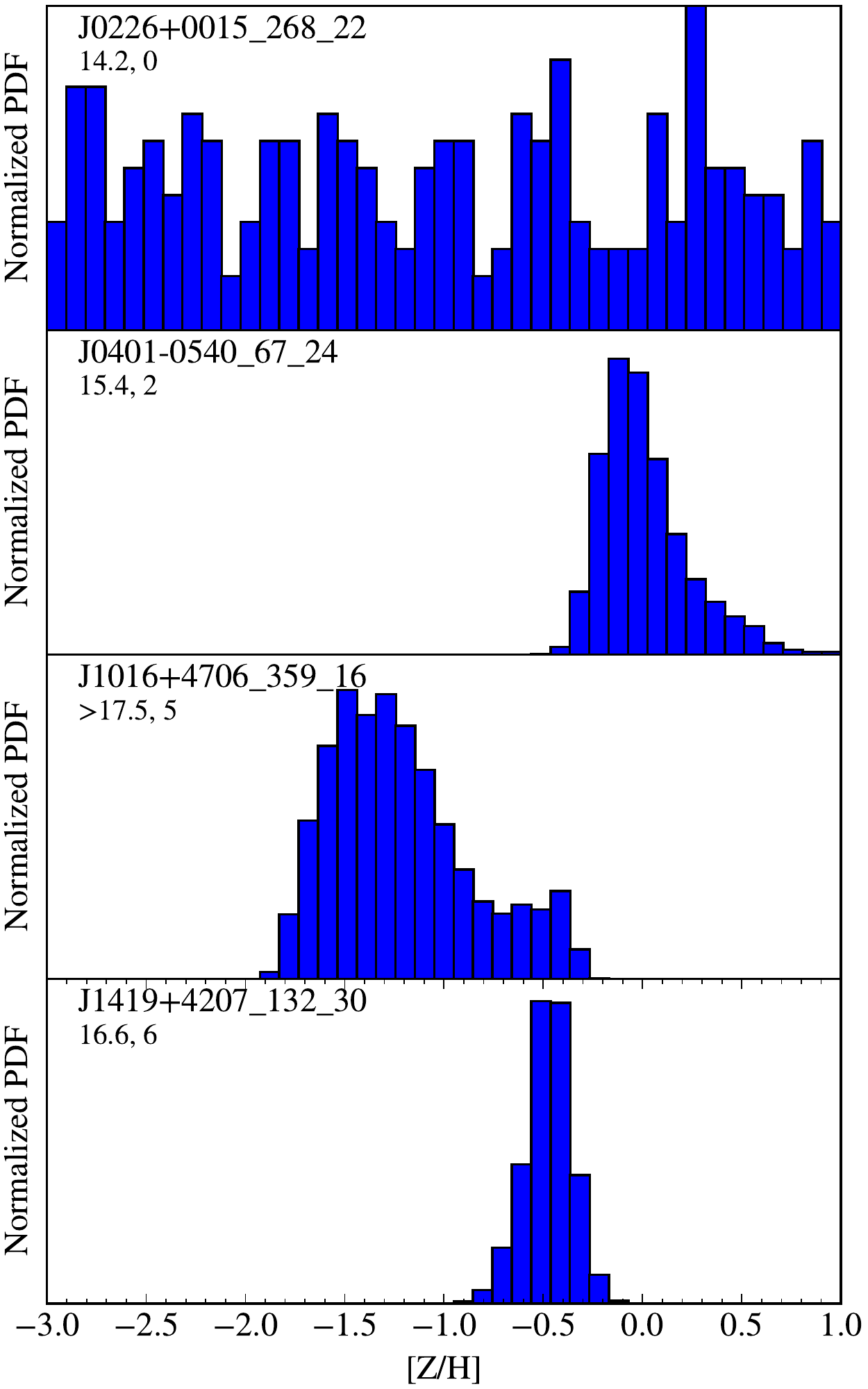}
\end{center}
\caption{
Derived metallicity PDFs for four representative
CGM systems from the COS-Halos sample. 
The top panel shows an example without a constraint on
any heavy element.  The second system (J0401-0540\_67\_24)
exhibits a positive detection of Si$^{++}$ and a low
\nhi\ value which drives the solution to a high [Z/H] value.
The last two systems exhibit many ionic transitions.
The relatively large uncertainty in [Z/H] for 
J1016+4706\_359\_16 reflects the large uncertainty in \nhi\
for this system owing to a fully saturated Lyman limit.
Labels under each system name give the $\log \mnhi$ value
followed by the number of positive detections used to 
constrain the metallicity PDF.
}
\label{fig:exPDF}
\end{figure}

The MCMC analysis yields metallicity PDFs for cool CGM gas under the assumption of photoionization equilibrium. Figure~\ref{fig:exPDF} shows four PDFs for systems with a varying set of observational constraints. The top system
(J0226+0015\_268\_22) has no positive detections of any metal transition and therefore no meaningful constraint on the PDF.  

The second example (J0401$-$0540\_67\_24 with $\mnhi = 10^{15.45} \cm{-2}$) 
shows only  a single
$\N{\rm Si^{++}}$ detection\footnote{And \ion{O}{6} absorption, but that higher ionization state is not modeled in this analysis.
See \cite{stern+16} for a model that adopts a density profile
to model a wider range of ionization states.} and several upper limits from non-detections. The metallicity PDF is driven to
high values because there is a maximal Si$^{++}$/H$^0$ ratio 
for photoionization models which establishes a lower limit to the gas metallicity. 
The final two examples in Figure~\ref{fig:exPDF} are systems with a large set of ion constraints.
One system J1016+4706\_359\_16 has
an imprecise \nhi\ measurement 
and a correspondingly
large uncertainty on [Z/H]. The other, J1419+4207\_132\_30, shows that metallicities can be estimated to $\pm 0.2$ dex in the best circumstances.

\section{Ionization corrections for Super-solar Gas}

In Section~\ref{sec:super_solar} we reported on 
several CGM systems with estimated metallicities of
solar or even super-solar abundances.  These results were derived
from our MCMC analysis of the \ion{H}{1} column density and the
observed set of metals.  Qualitatively, however, the requirement
of a high metallicity may be inferred simply from the single
observational constraint on the ratio of $\N{Si^{++}}$ to \nhi.

\begin{figure}
\begin{center}
\includegraphics[width=3.5in]{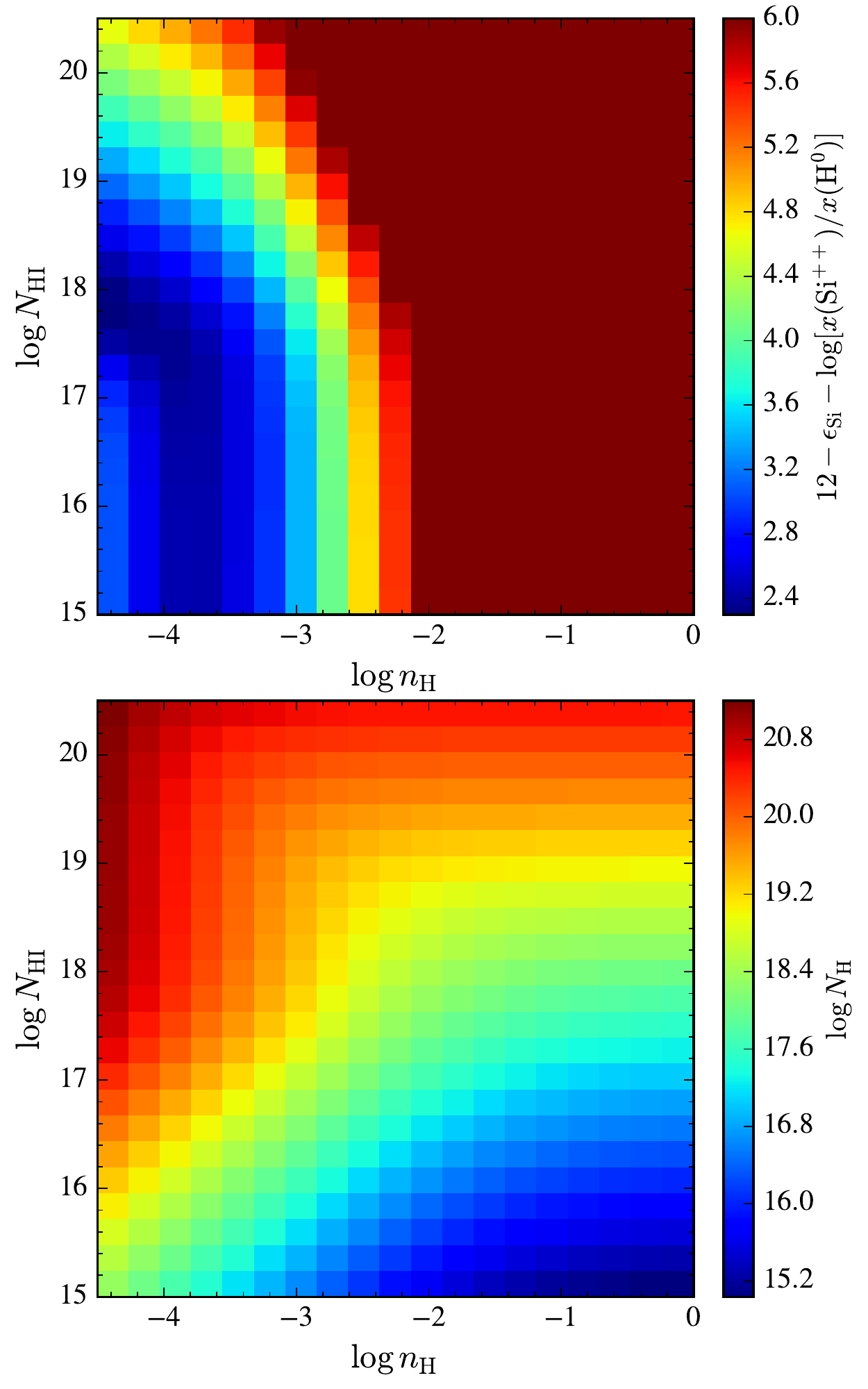}
\end{center}
\caption{
(top) Correction that must be applied to convert an 
observed $\log[\N{Si^{++}}/\mnhi]$ measurement to an
estimate of [Si/H] value for photoionization models
with the range of \nhi\ and $\log n_{\rm H}$ values
indicated. 
The minimum correction is $2.4$\,dex and this
sets a conservative lower limit to [Si/H] for a set
of systems in the COS-Halos sample.
(bottom) Derived total hydrogen column density 
$N_{\rm H}$ for the same models.
}
\label{fig:IC}
\end{figure}

Figure~\ref{fig:IC} presents the combined 
ionization and abundance corrections
required to convert an
observed $\log[\N{Si^{++}}/\mnhi]$ measurement 
to an estimate of [Si/H] value.   
The ionization corrections assume photoionization
equilibrium and a gas with solar metallicity
(adopting a lower metallicity would imply
a small difference in the calculation).
Examining the figure, one notes that the smallest
correction is $\approx 2.4$\,dex and occurs
for gas with low \nhi\ and low density 
(i.e.\ a high ionization parameter).
Therefore, under the assumption of photoionization equilibrum
(the results are similar for collisional ionization),
any system exhibiting
$\log [\N{Si^{++}} - \mnhi] > -2.4$\,dex indicates
at least a solar abundance of Si.

\end{document}